# CLIMATE-CONTINGENT FINANCE

*John J. Nay[1]*

## ABSTRACT

*Climate adaptation (reducing vulnerability to future climate change) could yield significant benefits. However, the uncertainty of which future climate scenarios will occur decreases the feasibility of proactively adapting. Fortunately, climate adaptation projects could be underwritten by benefits paid for in the climate scenarios that each adaptation project is designed to address because other entities would like to hedge the financial risk of those scenarios.*

*For instance, many infrastructure projects can be built to withstand extreme climate change through upfront spending. The climate adaptation expenditures generate more climate resilience benefit under more extreme climate. Because the return on investment of many adaptation actions is a function of the level of climate change, it is optimal for the adapting entity to finance adaptation with repayment that is also a function of the climate. It is also optimal for entities with more financial downside under a more extreme climate to serve as an investing counterparty because they can obtain higher than market rates of return when they need it most.*

*In this way, communities, cities, and states proactively adapting would reduce the risk they over-prepare, while their investors would reduce the risk they under-prepare. This is superior to typical insurance because, by investing in climate-contingent mechanisms, investors are not merely financially hedging but also outright preventing physical damage, and therefore creating economic value. Both sides of the positive-sum relationship — physical and financial hedgers — are made better off. This coordinates capital through time and place according to parties' risk reduction capabilities and financial profiles, while also providing a diversifying investment return to investors.*

*Governments, asset owners, and companies reduce uncertainty in components of the economy (e.g., commodities prices, credit risks, and interest rates) through trillions of dollars of derivatives positions and insurance contracts – this Article proposes a solution to provide a similar capability in the climate context. Municipalities raise trillions of dollars of debt for infrastructure – this Article proposes to provide that type of investment flow for financing climate-aware real asset projects.*

---

[1] Contact: jnay@nyu.edu. John holds a Ph.D. from Vanderbilt University and was previously an Adjunct Professor at NYU School of Law. He is currently the co-founder and CEO of Brooklyn Artificial Intelligence Research, and its wholly-owned subsidiary, Brooklyn Investment Group (bkln.com), which is registered as an Investment Adviser with the SEC. Nothing in this Article should be construed as financial or investment advice. This research represents his personal views and was not funded by Brooklyn Artificial Intelligence Research.

I would like to thank the following people for their input to this research: Oliver Schwartz, the editors of the *Berkeley Business Law Journal* (especially Nathaniel Whitthorne), William Warren, John Jacobi, Scott Worland, Hiba Baroud, J.B. Ruhl, Erkko Etula, Beth Richtman, Matthew Brand, Jim Gebhart, John Ryan, Moyo Ajayi, James Comfort, Hallie Hite, and Corey Hoffstein.

*Climate-contingent finance is a fresh approach to addressing catastrophic risk, building a bridge between long-term funding needs and financial risk management. It can be generalized to any situation where multiple entities share exposure to a risk where they lack direct control over whether it occurs (e.g., climate change, or a natural pandemic), and one type of entity can take proactive actions to benefit from addressing the effects of the risk if it occurs (e.g., through innovating on crops that would do well under extreme climate change or vaccination technology that could address particular viruses) with funding from another type of entity that seeks a targeted financial return to ameliorate the downside if the risk unfolds. This approach can finance previously under-funded efforts to address risks to humanity's long-term flourishing, including extreme climate change, large asteroids hitting the earth, and supervolcanic eruptions.*

TABLE OF CONTENTS

## I. INTRODUCTION

Reducing greenhouse gas emissions is a global collective action problem. Climate change adaptation (reducing vulnerability to future climate change) is a localized planning and financing problem with fewer generalizable solutions.[2] Although adaptation could yield significant benefits under future climate scenarios and cities will be spending trillions on infrastructure, the uncertainty of the climate scenarios decreases the feasibility of proactively adapting, especially where local political consensus is required (which is the case for most infrastructure in the U.S.). Due to human-caused climate change, uncertainty in the future climate is higher than it was when the processes for designing and financing infrastructure were developed. Furthermore, adapting to future climate scenarios is not just an issue for infrastructure designed explicitly for reducing climate risks, but is also increasingly a general problem for nearly all existing and future real assets. However, adaptation could be underwritten by benefits paid for in climate scenarios that the adaptation is designed to address because other entities would like to hedge the financial risk of those scenarios and support climate resilience.

For instance, infrastructure projects can be built to withstand or defend against extreme climate change through upfront spending. In doing so, infrastructure owners would be implicitly buying out-of-the-money options on more extreme change. These expenditures generate more climate resilience benefit under more extreme climate. When the return on investment of adaptation actions is a function of the level of climate change, it is optimal for the adapting entity to finance adaptation with repayment that is also a function of the climate. It is also optimal for entities with financial downside under a more extreme climate to serve as an investing counterparty because they can obtain higher than market rates of return when they need it most.

In this way, communities, cities and states proactively adapting would reduce the risk they *over*-prepare while their investors would reduce the risk they *under*-prepare. This is superior to typical insurance because, by investing in climate-contingent financial mechanisms, investors are not merely financially hedging but also outright helping prevent physical damage, and therefore creating economic value. Instead of buying insurance, they are paying for defense. Both sides of the positive-sum relationship — physical and financial hedgers — can be made better off. This coordinates capital through time and place according to parties' risk reduction capabilities and financial profiles, while also providing a diversifying investment return to investors.

Currently, governments, asset owners, companies, and farmers reduce uncertainty in components of the economy that impact their operations (e.g., commodities prices, credit risks, and interest rates) through trillions of dollars of derivatives positions[3] and insurance contracts – we propose to provide a similar capability in the climate context. Municipalities raise trillions of dollars of debt for infrastructure – we propose to provide that type of investment flow for financing climate-aware real asset projects.

---

[2] According to the World Bank, "when considering the total number of inventions across all technologies in all fields, the share of climate adaptation inventions in 2015 was roughly the same as in 1995. This stagnation of research and development for adaptation stands in sharp contrast to the trend for climate change mitigation technologies, whose share in total innovation [...] more than doubled during the same period." Matthieu Glachant, *Innovation in Climate Change Adaptation: Does it Reach Those Who Need it Most?*, WORLD BANK BLOGS (June 9, 2020), https://blogs.worldbank.org/climatechange/innovation-climate-change-adaptation-does-it-reach-those-who-need-it-most.

The remainder of this Article is set out as follows: in Part II, the Article describes the problem of climate uncertainty in long-term financing. Part III identifies the solution of climate-contingent instruments and explores the generalized structure of this mechanism. Such risk-contingent instruments apply to any situation where multiple entities share exposure to a catastrophic risk (e.g., climate change or a pandemic), and one type of entity can take proactive actions to benefit from addressing the risk if it occurs (e.g., through innovating on crops that would do well under extreme climate change or a vaccination technology that would address particular viruses) with funding from another type of entity that seeks a targeted financial return to ameliorate the downside if the risk unfolds. Part IV investigates the specifics of how climate-contingent financial instruments would work through case studies of three cities. Part V describes the types of potential participating parties to climate-contingent instruments. Part VI reports on extensive simulation experiments with climate-contingent contracts and optimization analyses of the simulation models. Part VII explores climate contingent bonds in more detail. Part VIII outlines a policy proposal to catalyze climate-contingent finance. Part IX concludes.

## II. PROBLEM: CLIMATE CHANGE & UNCERTAINTY

*"[Climate uncertainty] was marginal during previous centuries and, therefore, was often neglected in decision-making. Now, uncertainty in future climate change is so large that it makes many traditional approaches to designing infrastructure and other long-lived investments inadequate."*[4]

Climate change depends on many political, social, and environmental factors.[5] A sharp reduction in greenhouse gases would be ideal; however, the development of the political, social,

---

[3] For a sense of scale, consider that there are currently approximately $200 trillion just in derivatives contracts tied to LIBOR values. *See* Alex Harris & William Shaw, *Libor Proving Hard to Kill in $200 Trillion Derivatives Market*, BLOOMBERG (Jan. 11, 2021), https://www.bloomberg.com/news/articles/2021-01-11/libor-proving-hard-to-kill-in-200-trillion-derivatives-market.

[4] Stéphane Hallegatte, *Strategies to Adapt to an Uncertain Climate Change*, 19 GLOB. ENV'T CHANGE 240, 246 (May 2009), https://www.sciencedirect.com/science/article/pii/S0959378008001192.

[5] For instance, sea-level rise is a function of global emissions, the effect of emissions on temperature, and the effect of temperature on oceanographic changes, according to the *New York City Panel on Climate Change*, NYC MAYOR'S OFF. OF CLIMATE RESILIENCY, https://www1.nyc.gov/site/orr/challenges/nyc-panel-on-climate-change.page (last visited Dec. 26, 2021). For more on uncertainty in sea-level rise and its implications, see generally Robert E. Kopp et al., *Usable Science for Managing the Risks of Sea-Level Rise*, 7 EARTH'S FUTURE 1235 (2019), https://agupubs.onlinelibrary.wiley.com/doi/full/10.1029/2018EF001145. *See also* Marjolijn Haasnoot et al., *Generic Adaptation Pathways for Coastal Archetypes Under Uncertain Sea-Level Rise*, ENV'T RSCH. COMMC'NS 1 (2019), https://iopscience.iop.org/article/10.1088/2515-7620/ab1871 ("Adaptation to coastal flood risk is hampered by high uncertainty in the rate and magnitude of sea-level rise. Subsequently,

and technological solutions necessary remains uncertain.[6] For many scientific problems, uncertainty decreases with time. However, because of the positive feedback effects inherent in climate and socio-political systems,[7] uncertainty in the sensitivity of the climate to the level of emissions may *increase* over time. Uncertainty in the impact of carbon emissions on global temperature change has increased between the last release of the widely trusted global climate model ensemble and the release currently underway.[8] Fully accounting for the uncertainty in climate change significantly increases the costs of climate change and the expected benefits of adaptation.[9]

Figure 1 compares projections from the U.S. National Oceanic and Atmospheric Administration (NOAA) of the number of days per year of significant flooding under low (left) and extreme (right) climate scenarios for 99 coastal U.S. cities.

---

adaptation decisions carry strong risks of under- or over-investment, and could lead to costly retrofitting or unnecessary high margins.").

[6] *See generally* Hannah Nissan et al., *On the Use and Misuse of Climate Change Projections in International Development*, 10 WIREs Climate Change, no. 579, https://doi.org/10.1002/wcc.579; *see generally* Tanya Fiedler et al., *Business Risk and the Emergence of Climate Analytics*, 11 Nature Climate Change 87, 91 (2021), https://www.nature.com/articles/s41558-020-00984-6; *see generally* Peiran R. Liu & Adrian E. Raftery, *Country-Based Rate of Emissions Reductions Should Increase by 80% Beyond Nationally Determined Contributions to Meet the 2 ºC Target*, 2 Commc'ns Earth & Env't, no. 29, 1, 6-7, https://www.nature.com/articles/s43247-021-00097-8. Even if we know the level of emissions, "GCMs, although relatively consistent for global average results, exhibit large inter-model variability for regional climate projections." Lei Zhao et al., *Global Multi-Model Projections of Local Urban Climates*, 11 Nature Climate Change 152, 152 (2021), https://www.nature.com/articles/s41558-020-00958-8.

[7] *See, e.g.*, Zeke Hausfather & Richard Betts, *Analysis: How 'Carbon-Cycle Feedbacks' Could Make Global Warming Worse*, Carbon Brief (Apr. 14, 2020), https://www.carbonbrief.org/analysis-how-carbon-cycle-feedbacks-could-make-global-warming-worse.

[8] *See generally WCRP Coupled Model Intercomparison Project*, WCRP: World Climate Rsch. Programme (2021), https://www.wcrp-climate.org/wgcm-cmip; *see generally* Zeke Hausfather, *CMIP6: The Next Generation of Climate Models Explained*, Carbon Brief (Dec. 2, 2019), https://www.carbonbrief.org/cmip6-the-next-generation-of-climate-models-explained. Even the uncertainty in simulating historical temperature observations — in hindcasting (not forecasting) — has increased since the last release. For a comparison of the implications for sea-level rise, see generally Stefan Hofer et al., *Greater Greenland Ice Sheet Contribution to Global Sea Level Rise in CMIP6*, 11 Nature Commc'ns, no. 6289, Dec. 15, 2020, https://www.nature.com/articles/s41467-020-20011-8.

[9] *See generally* Raphael Calel et al., *Temperature Variability Implies Greater Economic Damages from Climate Change*, 11 Nature Commc'ns, no. 5029, Oct. 6, 2020, at 1, https://www.nature.com/articles/s41467-020-18797-8/.

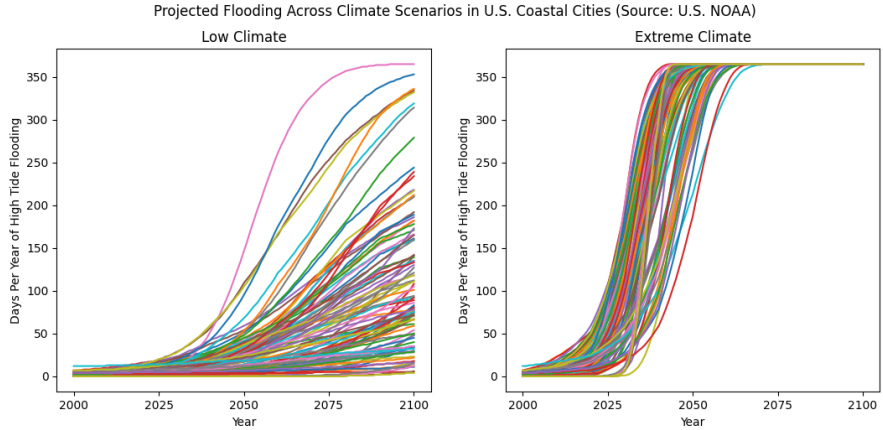

**Figure 1**: Yearly projections of days per year that exceed a significant flooding threshold, under low climate scenario (left) and extreme climate scenario (right) for 99 U.S. coastal cities. Each location is a separate line in each chart.[10]

In Figure 2, we focus on one city, New York, to visualize the difference between climate assumptions over varying time periods. The long tail in the changes in the extreme scenario for NYC is even longer when we move from a five-year to a ten-year horizon. Within some ten-year periods, projected increases of the number of days per year of significant flooding range from zero to more than 150, from *nothing* to *catastrophic*.[11]

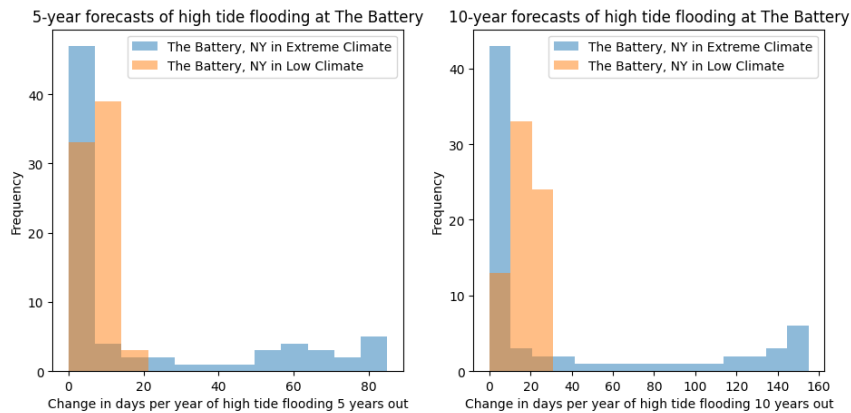

**Figure 2**: NOAA projections of days per year that exceed a significant flooding threshold for lower Manhattan in NYC from 2020-2100 transformed into change over *n*-years, where *n* is 5 or 10, for the low climate scenario (orange bars) and extreme climate scenario (blue bars). The extreme climate scenario has a fat right tail in its distribution.

---

[10] *See generally* data from WILLIAM V. SWEET ET AL., NOAA TECHNICAL REPORT NOS CO-OPS 096: PATTERNS AND PROJECTIONS OF HIGH TIDE FLOODING ALONG THE U.S. COASTLINE USING A COMMON IMPACT THRESHOLD 41-43 (Feb. 2018), https://tidesandcurrents.noaa.gov/publications/techrpt86_PaP_of_HTFlooding.pdf.

[11] The catastrophic outcomes would be due to passing "tipping points" in the climate. The consensus estimates of the temperature at which a tipping point could be reached continues to be moved lower as the science is better understood. Timothy M. Lenton et al., *Climate Tipping Points – Too Risky to Bet Against*, NATURE, Apr. 9, 2020, https://www.nature.com/articles/d41586-019-03595-0.

The ten-year forecasts (right chart) have a much more extreme right tail (note the x-axis range) than the five-year forecasts (left chart).[12]

Gavin Schmidt, one of the world's top climate scientists, said, "Do we have enough information to know that sea level is rising? Yes. Do we have enough information to tell people whether to build a 1-meter wall or a 2-meter wall? The answer is no."[13] Figure 3 supports his assertion, illustrating sea-level rise uncertainty across a wide range of possible climate scenarios.

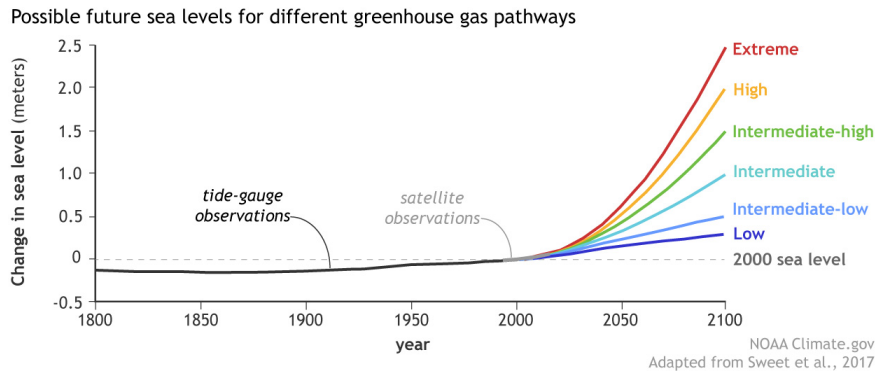

**Figure 3**: NOAA projections of sea level change.[14]

Due to human-caused climate change, uncertainty in the future climate is higher than it was when the processes for designing and financing infrastructure were developed. A recent $14 billion upgrade to New Orleans' flood infrastructure now looks like it could be inadequate in just four years because the designs underestimated sea-level rise.[15] Going forward, is the new design going to be an over- or under-estimate?[16] Extreme but scientifically plausible climate scenarios would be devastating if not proactively addressed. However, the high uncertainty over whether an extreme scenario will materialize — and therefore the potential to undertake an overprotective

---

[12] *See generally* data from WILLIAM V. SWEET ET AL., NOAA TECHNICAL REPORT NOS CO-OPS 096: PATTERNS AND PROJECTIONS OF HIGH TIDE FLOODING ALONG THE U.S. COASTLINE USING A COMMON IMPACT THRESHOLD 41-43 (Feb. 2018), https://tidesandcurrents.noaa.gov/publications/techrpt86_PaP_of_HTFlooding.pdf.

[13] Doug Struck, *Gavin Schmidt: The Problem with Climate Models? People.*, CHRISTIAN SCI. MONITOR (Jan. 22, 2021), https://www.csmonitor.com/Environment/2021/0122/Gavin-Schmidt-The-problem-with-climate-models-People.

[14] *See generally* data from WILLIAM V. SWEET ET AL., NOAA TECHNICAL REPORT NOS CO-OPS 096: PATTERNS AND PROJECTIONS OF HIGH TIDE FLOODING ALONG THE U.S. COASTLINE USING A COMMON IMPACT THRESHOLD (2018), https://tidesandcurrents.noaa.gov/publications/techrpt86_PaP_of_HTFlooding.pdf.

[15] Thomas Frank, *After a $14-Billion Upgrade, New Orleans' Levees Are Sinking*, SCI. AM. (2019), https://www.scientificamerican.com/article/after-a-14-billion-upgrade-new-orleans-levees-are-sinking/.

[16] Projections of sea-level rise estimated on climate model responses fall below simple extrapolation based on recent observational data, i.e., sea-level rise is even worse than the models thought it would be. *See generally* Aslak Grinsted & Jens Hesselbjerg Christensen, *The Transient Sensitivity of Sea Level Rise*, 17 OCEAN SCI. 181 (2021), https://os.copernicus.org/articles/17/181/2021/#:~:text=We%20define%20a%20new%20transient,temperature%20increases%20on%20this%20timescale.

project — can raise the cost of capital to prohibitive levels, or, more specifically, reduce the willingness to raise capital when the cost of that capital is not tied to the climate outcomes.[17] Louisiana's Coastal Protection and Restoration Authority developed a $50 billion plan to safeguard coastal populations, and only about $10 billion has been identified to support the plan.[18]

As a report from the Hoover Institution at Stanford University puts it: climate "uncertainty is new and distinct from risks that engineers routinely consider. It creates challenges for infrastructure planners and engineers unaccustomed to managing such ambiguities. There is a risk of over- or underbuilding, which can, in turn, transfer risks to infrastructure investors."[19] Most existing infrastructure is likely under-built, and some may be over-built, given the difficulty of making climate projections and embedding them into design processes.[20]

Goldman Sachs Global Markets Institute believes that:

> [A]daptation could drive one of the largest infrastructure build-outs in history . . . Given the scale of the task, urban adaptation will likely need to draw on innovative sources of financing. . . . Cities won't want to over-commit to specific climate scenarios. . . . Taking an investment approach might suggest that it makes sense instead to "wait and see," allowing time for new information to emerge before making any major investments. While this approach makes sense in many contexts, the case of climate change appears to be different. The most significant effects of climate change are likely to be the result of "tail events," which are inherently unpredictable in both their timing and their severity. Waiting won't necessarily generate more information about these idiosyncratic events. Waiting may instead mean that cities run out of time to prevent severe damages.[21]

An attractive approach to handling this uncertainty is to take adaptation actions that are designed to have payoffs (by preventing harms) that are as similar as possible across as many plausible climate outcomes as possible, so called "no-regrets" adaptation actions (e.g., building extreme weather early-warning systems). No-regrets actions should be pursued and can likely be financed by traditional mechanisms.[22]

---

[17] Regarding the cost of capital for climate risk investments, Pittsburgh Water & Sewer Authority said: "We are constantly evaluating ways to reduce our borrowing costs." Executive Director of Pittsburgh Water & Sewer Authority: "We need to look ahead and determine whether or not we should be thinking of a much higher degree of prevention or protection and how much can we afford." J. Dale Shoemaker, *How Pittsburgh is Funding the Fight Against Climate Change,* PUBLICSOURCE (Sept. 30, 2019), https://www.publicsource.org/how-pittsburgh-is-funding-the-fight-against-climate-change/.

[18] *Financing Resilient Communities and Coastlines: How Environmental Impact Bonds Can Accelerate Wetland Restoration in Louisiana and Beyond*, ENV'T DEF. FUND 16 (Aug. 2018), https://www.edf.org/sites/default/files/documents/EIB_Report_August2018.pdf.

[19] ALICE HILL ET AL., READY FOR TOMORROW: SEVEN STRATEGIES FOR CLIMATE-RESILIENT INFRASTRUCTURE 4 (Hoover Inst., 2019), https://www.hoover.org/research/ready-tomorrow-seven-strategies-climate-resilient-infrastructure.

[20] *See generally* B. Shane Underwood et al., *Past and Present Design Practices and Uncertainty in Climate Projections are Challenges for Designing Infrastructure to Future Conditions*, 26 J. INFRASTRUCTURE SYS. (2020) https://ascelibrary.org/doi/abs/10.1061/%28ASCE%29IS.1943-555X.0000567.

[21] Amanda Hindlian et al., *Taking the Heat: Making Cities Resilient to Climate Change*, GOLDMAN SACHS GLOB. MKT. INST., (Sept. 4, 2019), https://www.goldmansachs.com/insights/pages/gs-research/taking-the-heat/report.pdf.

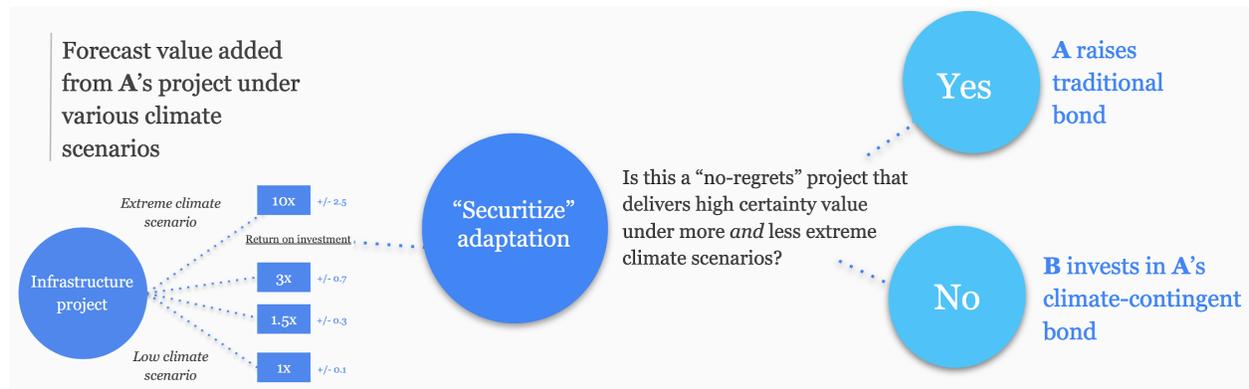

**Figure 4**: Climate-aware financial analysis of an infrastructure project.

However, truly no-regrets actions that significantly reduce physical risk are becoming rarer as more extreme scenarios are increasingly considered plausible. Some adaptation actions have to be designed and built, conditional on climate change scenarios. Yet it is difficult to design an action that would have similar payoffs across all plausible outcomes. An analysis of adaptation in the United States, the Philippines, and Britain concluded that, for the study areas, none of the flood protection projects had positive expected value for current climate conditions or a low climate change scenario, but all the strategies were economically attractive in the high climate change scenario.[23]

Few adaptation actions will deliver homogeneous benefits across climate scenarios.[24] Therefore, climate uncertainty translates into uncertainty in the benefits delivered by many adaptation actions. This is the core of the adaptation financing dilemma.

---

[22] Especially when adaption actions have general economic development co-benefits. *See generally* John J. Nay et al., *A Review of Decision-Support Models for Adaptation to Climate Change in the Context of Development*, 6 CLIMATE CHANGE 357 (Feb. 10, 2014), https://www.tandfonline.com/doi/full/10.1080/17565529.2014.912196. *See* Figure 4 for a visual explanation.

[23] Jeroen Aerts et al., *Evaluating Flood Resilience Strategies for Coastal Megacities*, 344 SCIENCE 473, 473-75 (May 2, 2014), https://www.science.org/doi/full/10.1126/science.1248222.

[24] S. Hallegatte et al., *Strengthening New Infrastructure Assets: A Cost-Benefit Analysis* 11 (World Bank Pol'y Rsch. Working Paper No. 8896, 2019), https://papers.ssrn.com/sol3/papers.cfm?abstract_id=3430506 ("[C]limate change makes the strengthening of infrastructure assets even more important. Without climate change, the median benefit-cost ratio would be equal to 2, but it is doubled when climate change is considered. And the fraction of scenarios in which strengthening infrastructure is not profitable is decreased from 14 to 4 percent when climate change is taken into account."); C.M. Shreve & I. Kelman, *Does Mitigation Save? Reviewing Cost-Benefit Analyses of Disaster Risk Reduction*, 10 INT'L J. DISASTER RISK REDUCTION 213, 231 (Dec. 2014), https://www.sciencedirect.com/science/article/pii/S2212420914000661 ("For changes in flood regimes, as a result of climate change as well as infrastructure development, understanding the hazard and vulnerability changes is much more challenging with larger uncertainties. DRR [disaster risk reduction] CBAs [cost-benefit analyses] might have different levels of usefulness depending on the hazard and depending on the hazard drivers, such as climate change, which are considered for analyzing CBAs in forward-looking studies"); *see*

Furthermore, adapting to future climate scenarios is not just an issue for infrastructure designed explicitly for reducing climate risks. It is a general problem for nearly all existing and future real assets. According to Morgan Stanley, "climate resilience is fast becoming an investment imperative in real assets."[25] Owners of real assets (infrastructure, buildings, and land) have long investment holding periods (often decades)[26] and high exposure to climate change impacts. However, they currently have no means to hedge this long-term climate uncertainty.[27] Hannah Nissan et al. point out the incongruity in how climate uncertainty is treated compared to other complex systems:

> Foresight about future exchange rates, oil prices, geopolitical disruptions, or epidemics of new diseases would be invaluable, but there is little expectation that such things can accurately be forecast beyond the short term. Despite high confidence in many aspects of present and future climate change, localized projections are highly unreliable. Where then does the unrealistic expectation come from that the future climate, among the most complex of known systems, should be predictable to the degree of precision often demanded?[28]

Long-term financial entities owning real assets will be forced into one of two groups: Adapters ("*A*"), proactively adapting; or Backers ("*B*"), absorbing impacts.[29] Both groups face obstacles that could be addressed by collaborating through a financial mechanism. Some entities are better positioned to move into Group *A* and reduce their physical risk now; while others will determine that reducing their financial risk without immediately reducing their physical risk is more feasible, and move into Group *B* for the time being.[30]

---

*generally* Borja G. Reguero et al., *Comparing the Cost Effectiveness of Nature-Based and Coastal Adaptation: A Case Study from the Gulf Coast of the United States*, 13 PLOSONE 1, 8 (Apr. 11, 2018), https://journals.plos.org/plosone/article?id=10.1371/journal.pone.0192132; *see generally* Audrey Baills et al., *Assessment of Selected Climate Change Adaptation Measures for Coastal Areas*, 185 OCEAN & COASTAL MGMT, Mar. 1, 2020, at 1, https://www.sciencedirect.com/science/article/pii/S0964569119309287.

[25] MORGAN STANLEY, WEATHERING THE STORM: INTEGRATING CLIMATE RESILIENCE INTO REAL ASSETS INVESTING 3 (2018), https://www.morganstanley.com/im/publication/insights/investment-insights/ii_weatheringthestorm_us.pdf. Adaptation actions may have been taken during previous climates as well; L. Supriya, *Taíno Stilt Houses May Have Been an Adaptation to Climate Change*, EOS (Jan. 15, 2021), https://eos.org/articles/taino-stilt-houses-may-have-been-an-adaptation-to-climate-change.

[26] E.g., coal-fired power plants are designed for 40 to 50 years of production, hydropower infrastructure is designed for up to 100 years, and approximately 66% of U.S. city infrastructure is more than 30 years old today. Jonathan Woetzel et al., *Will Infrastructure Bend or Break Under Climate Stress?*, MCKINSEY GLOB. INST. (Aug. 19, 2020), https://www.mckinsey.com/business-functions/sustainability/our-insights/will-infrastructure-bend-or-break-under-climate-stress.

[27] See *infra* Appendix A: Climate Risk Pricing for a discussion of how climate uncertainty affects asset pricing and why real asset owners are the most vulnerable to physical climate-induced price changes.

[28] Nissan et al., *supra* note 6, at 6.

[29] E.g., a city, a homeowner, a mortgage lender, a state government or a transportation group such as the New York City Metropolitan Transportation Authority that runs the public transportation system.

[30] An entity could be an *A* or a *B* at different time scales and geographies. There will likely be more *B* parties than *A* parties, i.e., a lot of entities would like to insure at least a small amount of their climate risks but not as many can build large physical adaptation projects.

Currently, most **B** entities are unable to effectively hedge the risk of climate outcomes. Municipal bonds have long-term climate risk. Given that there is no way to cleanly hedge municipal bond credit risk (the primary risk to these securities),[31] there is even less opportunity to cleanly hedge municipal bond climate risk. The vast majority of insurance is on a one-year time horizon, which is not helpful for locking in certainty of a hedge on a time scale relevant to climate change, because every year insurance providers can increase rates or stop providing insurance altogether.[32] Furthermore, insurance does not reduce risk in the aggregate. As a "risk transfer" mechanism, it merely shifts risk from one party to another.[33] If we physically reduce risk and prevent damages, we can generate more overall value and, in effect, share that value between parties. Therefore, a standard parametric insurance payout would likely provide a lower expected return than a triggered climate contract and serve as a less effective hedge.[34]

## III. SOLUTION: FINANCING ADAPTATION BY REDUCING COUNTERPARTIES' RISKS

Party **A** proactively adapts, making physical changes that explicitly take climate change scenarios into consideration, funded by **B**[35] hedging financial risk of climate-induced losses (Figure 5).[36] **B** provides upfront capital to **A**, who uses the proceeds for adaptation that will substantially reduce losses in more extreme climate scenarios. Under less extreme scenarios, the adaptation may be overprotective. A climate-related financial product of this nature was first proposed by Daniel Bloch and co-authors as a "climate default swap."[37]

---

[31] Ming-Jie Wang, *Credit Default Swaps on Municipal Bonds: A Double-Edged Sword?*, 35 YALE J. REG 301, 307 (2018).

[32] *See, e.g.*, Ange Lavoipierre & Stephen Smiley, *Could Climate Change Make it Harder to Get Insurance in Australia?*, THE SIGNAL NEWS (Feb. 5, 2019), https://www.abc.net.au/news/2019-02-06/could-climate-change-make-australia-uninsurable/10783490.

[33] "While current ILS instruments are useful for transferring risks, they are not designed to reduce underlying risks or build resilience to disasters." Lauren Carter, *Can Insurance-Linked Securities Mobilize Investment in Climate Adaptation?*, UNDP (Jan. 12, 2021), https://www.undp.org/blog/can-insurance-linked-securities-mobilize-investment-climate-adaptation.

[34] "As [parametric insurance policies have been taken out by municipalities,] those policies have gotten more popular, they've started to run into serious questions—like [...] whether the policies' existence allows governments to punt on harder decisions about where people live and businesses operate in the first place." Zack Colman, *Insurance for When FEMA Fails*, POLITICO (Jul. 14, 2020), https://www.politico.com/news/agenda/2020/07/14/climate-change-fema-insurance-341816.

[35] The plan is to have many **B** entities per **A**, with a fund that has **B**s as limited partners and invests in many climate contracts.

[36] *See generally* Daniel Bloch et al., *Applying Climate Derivatives to Flood Risk Management*, 56 WILMOTT 88 (2012), https://onlinelibrary.wiley.com/doi/abs/10.1002/wilm.10058 (discussing the concept of climate change derivatives to mitigate flood risk); *see also* Daniel Bloch et al., *Cracking the Climate Change Conundrum with Derivatives*, 2 WILMOTT J. 271 (2010), https://onlinelibrary.wiley.com/doi/abs/10.1002/wilj.41 (original article discussing the concept abstractly); *see* Figure 5 for a visual representation of climate derivatives.

[37] Bloch et al. (2010), *supra* note 36, at 271-272.

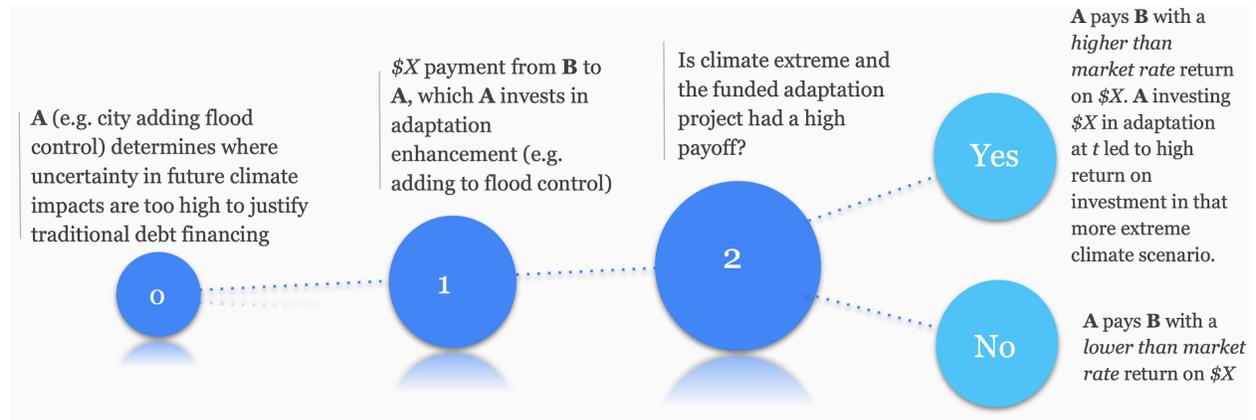

**Figure 5**: Climate contract lifecycle.

If the effects of climate change are worse than expected, *A* pays *B* back with a higher-than-market rate of return on the principal.[38] Reducing risk of property and human health damages is likely to reduce losses (or even increase benefits) in more extreme climate change scenarios.[39] In addition to averting damage, adaptation measures reduce *A*'s cost of capital for general operations by increasing their creditworthiness: according to BlackRock research, "bonds issued by climate-resilient states and cities are likely to trade at a premium to those of vulnerable ones over time."[40] *A* is insuring against risk of ruin by building adaptation projects and only paying for that "insurance" in scenarios where those projects are most needed. The financing allows *A* to take steps to realize the benefits of adaptation and hedge against overprotecting while doing so.[41]

---

[38] Daniel Bloch, James Annan, & Justin Bowles, *Cracking the Climate Change Conundrum with Derivatives*, 2 WILMOTT J. 271 (2010), https://onlinelibrary.wiley.com/doi/abs/10.1002/wilj.41.

[39] *NIBS Finds Investment in Resilient Design Can Pay Off by More than Sixfold*, ARCHITECT MAGAZINE (Jan. 18, 2018), https://www.architectmagazine.com/technology/nibs-finds-investment-in-resilient-design-can-pay-off-by-more-than-sixfold_o. According to Seung Kyum Kim, raising foundations provides 6.6% and 14.3% housing price increases in Miami-Dade and NYC, and adaptation for storm surges provides a 15.8% housing price increase in Miami-Dade. There is direct loss mitigation value in the event of a climate threshold crossing and, regardless of risk events, there is asset price appreciation due to the recognition of that resilience. Seung Kyum Kim, The Economic Effects of Climate Change Adaptation Measures: Evidence from Miami-Dade County and New York City 2, 24 (May 2019) (Doctor of Design dissertation fellowship working paper) (on file with the Joint Center for Housing Studies of Harvard University), https://www.jchs.harvard.edu/research-areas/working-papers/economic-effects-climate-change-adaptation-measures-evidence-miami; *see also* Delavane B. Diaz, *Estimating Global Damages from Sea Level Rise with the Coastal Impact and Adaptation Model (CIAM)*, 137 CLIMATIC CHANGE. 143, 143 (2016) ("[T]here is large potential for coastal adaptation to reduce the expected impacts of SLR compared to the alternative of no adaptation, lowering global net present costs through 2100 by a factor of seven . . . ."); *see also* HALLEGATTE ET AL., THE WORLD BANK, LIFELINES: THE RESILIENT INFRASTRUCTURE OPPORTUNITY 2 (2019), https://documents1.worldbank.org/curated/en/775891600098079887/pdf/Lifelines-The-Resilient-Infrastructure-Opportunity.pdf ("In the median [climate and economic] scenario, the net benefit of investing in more resilient infrastructure in low- and middle-income countries is $4.2 trillion, with $4 in benefit for each $1 invested. Climate change makes action on resilience even more necessary and attractive: on average, it doubles the net benefits from resilience. And because large investments in infrastructure are currently being made in low- and middle-income countries, the median cost of one decade of inaction is $1 trillion."). For an Intermediate-High NOAA climate scenario for Miami-Dade County in Florida, the return on investment for building-level adaptation actions is estimated to be 518%, and the return on investment for community-wide adaptation actions is estimated to be 926%. URBAN LAND INST. , RESEARCH REPORT: THE BUSINESS CASE FOR RESILIENCE IN SOUTHEAST FLORIDA: REGIONAL ECONOMIC BENEFITS OF CLIMATE ADAPTATION 4, 26 (2020), https://knowledge.uli.org/reports/research-reports/2020/the-business-case-for-resilience-in-southeast-florida?_gl=1*17yztlq*_ga*MjA2MDgwNTAwOC4xNjQwODk1MjE1*_ga_HB94BQ21DS*MTY0MDg5NTIxNC4xLjAuMTY0MDg5NTIxNC4w. A study of Ho Chi Minh City adaptation to sea-level rise finds very large spread in the benefit-cost ratios and net-present values of adaptation measures between medium and high climate change scenarios, Paolo Scussolini et al., *Adaptation to Sea Level Rise: A Multidisciplinary Analysis for*

Benefits of climate adaptation for a municipality include reduced insurance premiums,[42] reduced future uninsured direct damages to property and infrastructure assets, reduced future

---

*Ho Chi Minh City, Vietnam*, 53 WATER RES. RSCH. 10841, 10852 (2017), https://agupubs.onlinelibrary.wiley.com/doi/full/10.1002/2017WR021344 ("The combination of elevation + dryproofing shows the highest B/C [benefit-cost ratio], ranging from 41 in [the most probable possible future greenhouse gas concentrations (RCP4.5)] to 97 in [extremely severe possible future greenhouse gas concentrations (RCP8.5)] High-end, and the highest NPV, from 514 B$ in RCP8.5 High-end to 216 B$ in RCP4.5."); s*ee also* Hallegatte et al., *supra* note 24, at 11 ("[C]limate change makes the strengthening of infrastructure assets even more important. Without climate change, the median benefit-cost ratio would be equal to 2, but it is doubled when climate change is considered. And the fraction of scenarios in which strengthening infrastructure is not profitable is decreased from 14 to 4 percent when climate change is taken into account."); *see also* Reguero et al., *supra* note 24, at 15 ("As sea level rises, land subsides, storms increase in frequency and intensity, and assets in the coastal zone increase, all adaptation measures become more cost-effective . . . ."); *see also* Jeremy Martinich & Allison Crimmins, *Climate Damages and Adaptation Potential Across Diverse Sectors of the United States*, 9 NATURE CLIMATE CHANGE 397, 401 (2019) ("Projected physical and economic damages are larger under [extremely severe possible future greenhouse gas concentrations (RCP8.5)] than under [the most probable possible future greenhouse gas concentrations (RCP4.5)] across all 22 sectors and both time periods, with only 1 exception (urban drainage adaptation costs in 2050. . . ). Damages associated with extreme weather, such as extreme temperature, heavy precipitation, drought and storm surge events, are substantially reduced under RCP4.5. For example, more than twice as many 100-year riverine inland flooding events are projected across the CONUS under RCP8.5 compared to RCP4.5 by the end of the century . . . .").

[40] Ashley Schulten et al., *Getting Physical: Scenario Analysis for Assessing Climate-Related Risks*, BLACKROCK INV. INST. 1, 12, (2019), https://www.blackrock.com/ch/individual/en/insights/physical-climate-risks.

[41] *Bs*, and any financial intermediary facilitating the transaction, has an implicit incentive to help *As* ensure their adaptation investments are properly implemented because that increases the probability that the *As* will be able to pay the *Bs* back.

[42] The U.S. National Flood Insurance Program Community Rating System reduces premiums to reflect the reduced flood risk resulting from a community's efforts. *National Flood Insurance Program Community Rating System*, FEMA (last accessed Nov. 16, 2021), https://www.fema.gov/floodplain-management/community-rating-system?web=1&wdLOR=cCB3563C1-9027-8541-BD4B-80EEA92A7C56.

costs for rebuilding,[43] reduced potential litigation costs,[44] reduced cost of capital for borrowing,[45] maintained attractiveness of the area for outside investment and in-migration, maintained property tax, sales tax, and tourism tax revenues,[46] avoided tail-risk scenarios of collapsing property values and business activity that could lead to a downward spiral and complete abandonment, and increased revenue for natural adaptation solutions from selling carbon credits.[47]

Meanwhile, ***B*** is better off because its returns through its hedge are greater than returns from other investments in a more extreme climate outcome state of the world.[48] We explore values of the repayment rates in the simulation experiments below.[49] The rate of return required by ***B*** may be relatively low because the investment pays off specifically in states of the world with high marginal utility; in more extreme climate scenarios, a dollar is worth more than in less extreme climate scenarios.[50] Investing allows ***B*** to hedge against physical under-preparedness in a way that is directly linked to their climate exposures. ***B*** parties that are taking a "wait-and-see" approach to climate adaptation through participation in this investment can gain information on adaptation project outcomes — and generate capital in the triggered scenarios — to implement their own (less proactive) adaptation projects in the future.

---

[43] Additionally, while less of a direct monetary benefit than the list in the main text, climate change damages and the reconstruction they require can create greenhouse gas emissions. In this way, by reducing damages and the need to rebuild infrastructure, adaptation can reduce greenhouse gas emissions, which has monetary benefits, especially with a potential price placed on carbon emissions. More broadly, there can be emissions mitigation co-benefits to adaptation actions. *See generally* Lobell et al., *Climate adaptation as mitigation: the case of agricultural investments*, ENVIRONMENTAL RESEARCH LETTERS 8 (2013), https://iopscience.iop.org/article/10.1088/1748-9326/8/1/015012/meta.

[44] "[L]itigation could be packaged as breaches of duty, ordinary negligence, and inverse condemnation actions based on a public entity's failure to adequately plan, prepare, and invest for the inevitable effects of climate change. Any resulting unplanned expenditures due to climate change could potentially cause an inability of the municipality to honor their debt obligations [...] Plaintiffs have alleged that defendants (utilities in some cases) are guilty of wrongful acts or negligence because the defendants gave climate change impacts insufficient consideration, planning, and investment. For example, this occurred in cases related to the 2018 California wildfires, where insurance companies, citing inverse condemnation, are looking to PG&E to pay for wildfire losses." *Insurance, Bond Ratings and Climate Risk: A Primer for Water Utilities*, ASS'N MET. WATER AGENCIES 3, 6 (2019), https://www.amwa.net/assets/Insurance-BondRatings-ClimateRisk-Paper.pdf.
And for electric utilities, see Romany M. Webb, Michael Panfil, & Sarah Ladin, *Climate Risk in the Electricity Sector: Legal Obligations to Advance Climate Resilience Planning by Electric Utilities*, COLUMBIA L. SCH. SABIN CTR. CLIMATE CHANGE L. (Dec. 3, 2020), https://climate.law.columbia.edu/content/climate-risk-electricity-sector-legal-obligations-advance-climate-resilience-planning.

[45] Debt service costs can account for 50% of a water utility's total costs. Chapter 2 of Jeff Hughes et al., *Defining a Resilient Business Model for Water Utilities*, WATER RSCH. FOUND. (2014), https://www.researchgate.net/publication/277477105_Defining_a_Resilient_Business_Model_for_Water_Utilities. Those costs are higher if the credit ratings of the utilities are lower.
Pittsburgh Water & Sewer Authority said: "We are constantly evaluating ways to reduce our borrowing costs." Executive Director of Pittsburgh Water & Sewer Authority: "We need to look ahead and determine whether or not we should be thinking of a much higher degree of prevention or protection and how much can we afford." J. Dale Shoemaker, *How Pittsburgh is Funding the Fight Against Climate Change,* PUBLICSOURCE (Sept. 30, 2019), https://www.publicsource.org/how-pittsburgh-is-funding-the-fight-against-climate-change/.

[46] Maintaining, or even increasing, the market value of the privately held property could make a material difference to the property tax collected.

[47] For examples of adaptation related activities that could also have emissions reductions benefits, *see generally* Joseph E. Fargione et al., *Natural Climate Solutions for the United States*, 4 SCIENCE ADVANCES, 1, 1(2018)., https://www.science.org/doi/10.1126/sciadv.aat1869.

[48] If an ***A*** is a state or city government, it is very unlikely, and for some, illegal, to declare bankruptcy and avoid payments. However, the money needs to come from somewhere. The ***A*** will need to raise general real estate and sales taxes if they are a city or state, increase fees if they are a water or energy utility, or implement taxes that are closer to directly capturing the benefits provided by the adaptation project. *The Virtues of Value Capture*, DELOITTE (2019), https://www2.deloitte.com/content/dam/Deloitte/global/Documents/Public-Sector/smart-cities-virtues-of-value-capture-19nov.pdf, e.g., taxes on real estate nearest a seawall investment or beach nourishment; *see generally* Megan Mullin et al., *Paying to Save the Beach: Effects of Local Finance Decisions on Coastal Management*, CLIMATIC CHANGE, 275, 275 (2018), https://link.springer.com/article/10.1007/s10584-018-2191-5.

---

Arpit Gupta discusses how infrastructure projects could be funded by more specific property taxes: "Our paper demonstrates that it is technically feasible to determine how much each housing unit benefited from the new transit infrastructure, taking into account its exact location, and its unit and building characteristics. In theory, local government could levy a unit-specific property tax surcharge proportional to the value created. Such micro-targeted property tax surcharges would not only be based on objectively measurable value increases and property characteristics, and hence be fair, they could also become an important financing tool to fund future infrastructure needs." Arpit Gupta & Stijn Van Nieuwerburg, *Take the Q Train: Value Capture of Public Infrastructure Projects* 1, 39 (Nat'l Bureau of Econ. Rsch. Working Paper no. 26789, 2020), https://www.nber.org/papers/w26789.

"The value of a property protected by storm surge by a new seawall or natural barrier typically should be higher than the value of a similarly situated property that does not have such protection. Authorities can estimate the value difference between comparable properties with protection and those without it. Some of the difference can then be 'captured' through increased taxes on the benefitting properties." ALICE HILL & LEONARDO MARTINEZ-DIAS, BUILDING A RESILIENT TOMORROW: HOW TO PREPARE FOR THE COMING CLIMATE DISRUPTION 86 (Oxford Univ. Press 2020).

Furthermore, ***Bs*** can better make payments if ***As*** also issues a catastrophe (cat) bond. We could issue a cat bond linked to the same climate threshold in order to be guaranteed to be able to pay some portion of the outcome to ***Bs*** if the threshold is reached, reducing their counterparty risk. We would be issuing a given cat bond that covers many climate contracts, and there would be many ***B*** parties to a single contract, so we would be able to spread cat bond transaction costs across enough ***B*** parties that this would be a more attractive option to a ***B*** than issuing a cat bond themselves. Cat bonds have characteristics that make them well suited for backstopping or complementing a climate contract: they are 100% collateralized and so have essentially no counterparty risk, they provide time scales (multi-year) that lock in rates, and because they have more investor types than most insurance related instruments, they can more buyer demand and thus have reduced rates.

It is also worth noting that if there are acute disasters occurring, the federal government will likely be providing financial disaster aid after acute climate events.

[49] For an alternative pricing approach, see Bloch et al. (2012), *supra* note 36, at 89 (applying the logic of pricing credit derivative products to pricing climate derivatives by replacing the survival probabilities and default time densities with first-passage (of a climate threshold) distributions and first-passage time density).

[50] Risks that cannot be diversified away (systemic risks) are those that increase the probability that an asset's value is correlated with most other global asset values, see generally ANTTI ILMANEN, EXPECTED RETURNS: AN INVESTOR'S GUIDE TO HARVESTING MARKET REWARDS (2011); correlated with equity market volatility, see generally TIM LEE ET AL., THE RISE OF CARRY: THE DANGEROUS CONSEQUENCES OF VOLATILITY SUPPRESSION AND THE NEW FINANCIAL ORDER OF DECAYING GROWTH AND RECURRING CRISIS (2019). This is likely the case for the more extreme global climate risks; therefore, the marginal utility of a dollar is higher for the payoffs for strategies that hedge these risks. *See generally* Stefano Giglio et al., *Climate Change and Long-Run Discount Rates: Evidence from Real Estate* (Nat'l. Bureau Econ. Rsch. Working Paper no. 21767, 2015), https://www.nber.org/papers/w21767 at page 6. The Financial Stability Board believes that climate risk "may change – and in places, increase – the degree of co-movement between asset prices, and reduce the degree to which financial market participants were able to diversify exposure to climate-related risks. It might also reduce the efficacy of other channels through which financial market participants seek to insure against climate risks (e.g., via some derivatives markets)." THE IMPLICATIONS OF CLIMATE CHANGE FOR FINANCIAL STABILITY 17 (Fin. Stability Bd. Nov. 23, 2020), https://www.fsb.org/2020/11/the-implications-of-climate-change-for-financial-stability/.

If the effects of climate change are less severe than expected, ***A*** is no worse off than they would have been otherwise, and probably better off. First, they may have over-prepared (at least within the timeline of the repayment, potentially not later), but they paid less for it than a traditional bond repayment. Second, they are more prepared for a future increase in climate change that may still occur over a longer time period beyond the end of the repayment. Third, there are sometimes resilience "co-benefits" that the adaptation projects serve. ***B*** was repaid less than they would have been with a traditional debt investment, but their climate risk was hedged enough during the ensuing period that they were able to continue operating and borrowing at lower rates.

## A. *Generalized Structure*

This Article focuses on climate change, but we believe the generalized structure of this risk-contingent financing mechanism applies to any situation where multiple entities share exposure to a risk out of their direct control,[51] ***R***, and one type of entity, ***A***, can take proactive actions to benefit from (either through avoided losses or through absolute gains) addressing ***R*** if it occurs with funding from another type of entity, ***B***, that seeks a targeted financial return to ameliorate the downside if ***R*** unfolds.

Examples of risks, ***R***, that are appropriate for this type of financing include extreme climate change, natural pandemics, and large asteroids hitting the earth. Examples of proactive actions to benefit from addressing systemic risks include innovating on crops that would do well under extreme climate change, vaccination technologies that would address particular viruses, and mechanisms that deflect large asteroids from earth impact. The actions to mitigate ***R*** yield a payoff ***P*** if ***R*** occurs (through avoided losses or through absolute gains).

- It's difficult to take proactive action to a reduce risk (**R**) when there's an uncertain chance **R** will materialize
- If **R** does materialize, proactive action could have a massive payoff (**P**)
- Out of the entities with exposure to **R**: some can take proactive action (**A**) and others fund them (**B**)
- If **R** happens, **A** uses part of **P** to pay **B** back more than they would if **R** does not happen. This helps **B** with a targeted financial return to ameliorate their downside
- If **R** does not happen, **P** does not occur, but **A** and **B** were insured/hedged and were able to operate with risks managed

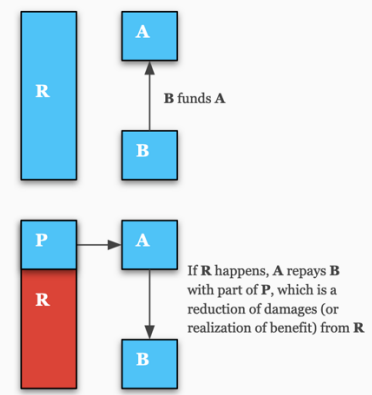

**Figure 6:** Generalized risk-contingent financing structure.

---

[51] The risk-contingent financing mechanism should be applied to situations where the parties involved have negligible direct control over whether the risk occurs. If the investor in the risk-contingent instrument could influence the likelihood of the risk occurring, it could lead to misaligned incentives because they may later on decide that the higher return received from the payout of the contract outweighs the downsides of the risk occurring. With large-scale risks such as extreme climate change, natural pandemics, or asteroids hitting the earth, there are no entities with material direct control over whether the risk is realized.

Insurance does not reduce risk, when measured in the aggregate; insurance shifts risk from one party to another.[52] If, instead, we physically reduce risk and prevent damages, we generate more overall value, which can be shared between parties. Therefore, a parametric insurance payout provides a lower expected return than a triggered contingent contract and serves as a less effective hedge.[53] The key is recognizing the payoffs that proactive risk reduction would have under the negative states of the world, and then, in effect, "securitizing" those payoffs to raise capital to fund the risk reduction.

Another important downside to traditional insurance is that it works by diversifying lowly correlated risks. If a source of risk is systemic, affecting most parties and therefore creating set of highly correlated risks, then it cannot be diversified away. To address systemic risk, it would need to be proactively reduced.

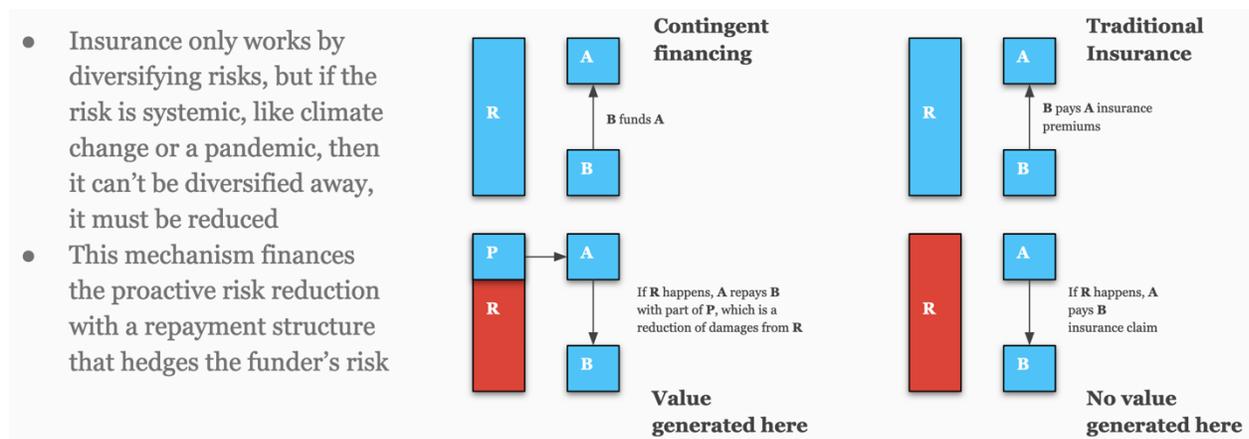

**Figure 7:** Comparison of risk-contingent financing to traditional insurance.

*A* makes changes that explicitly aim to reduce risk, funded by *B*[54] hedging financial risk. *B* provides upfront capital to *A*, who uses the proceeds to reduce their losses in more extreme negative scenarios (Figure 8). In the climate example, *A* might be building a tall seawall designed for extreme climate change, for instance. Under less extreme scenarios, the actions taken may be overprotective.

---

[52] "While current ILS instruments are useful for transferring risks, they are not designed to reduce underlying risks or build resilience to disasters." Carter, *supra* note 33.
[53] *See* Figure 7.
[54] Likely many *B* entities per *A*.

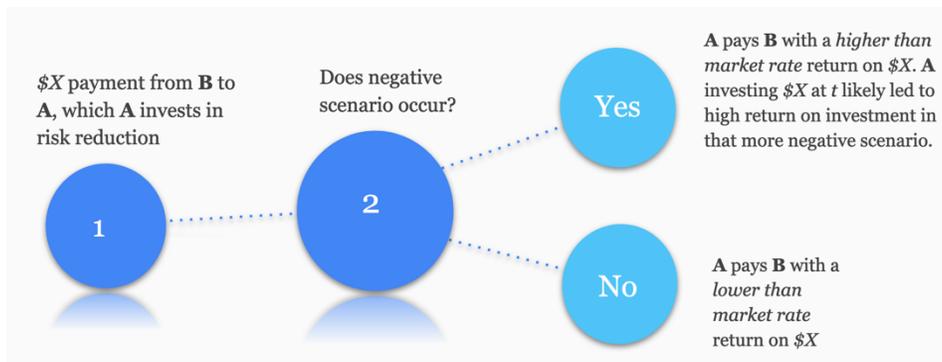

**Figure 8:** The generalized structure of the risk-contingent financing mechanism, which applies to any situation where multiple entities share exposure to an underlying systemic risk, and one type of entity, *A*, can take proactive actions to benefit from (either through avoided losses or through absolute gains) addressing that negative risk (if it occurs) with funding from another type of entity, *B*, that seeks a targeted financial return to ameliorate the downside (if it occurs).

The risk-contingent financing mechanism provides capital from *B* to *A* at issuance in return for the obligation that *A* pay back principal and a return if the negative scenario, or one more extreme, is realized before expiration of the contract, i.e., if *R* occurs in the specified time range. The amount *B* pays *A* initially (*Principal*), the amount that *A* would pay back *B* if triggered (*Return*), the *Scenario* beyond which triggers the payback, and the length of the *Term* within which the trigger must be passed to cause payout are all specified when the contract is initially sold.

Contract specifications at the time of initialization:
- *Principal* (e.g., $15 million)
- *Return* (e.g., 150%)
- *Scenario* (e.g., sea-level 1.5 inches above baseline for more than 1 year)
- *Term* (e.g., 15 years)

Contract participants:
- *A* (e.g., an airport building a seawall)
- *B* (e.g., a set of banks and insurance companies)

This general construct of a risk-contingent financial "contract" can be used to create single-trigger swap financial products at one time horizon (what we have described thus far), or debt-like products that have variable periodic interest rates contingent on the Scenario at multiple time horizons. The latter can be created by simply composing a multi-period repayment structure from a series of these contracts at different time horizons.

In the climate context, there is a spectrum spanning the extent to which repayment of principal is tied to a climate change outcome, with traditional debt at 0%, and the structure as described above at 100%. In between the two extremes, repayment could be partially tied to the climate variable; as the climate variable approaches the threshold, the repayment rate increases to a rate similar to traditional debt, and then surpasses that rate as the climate variable passes the threshold.

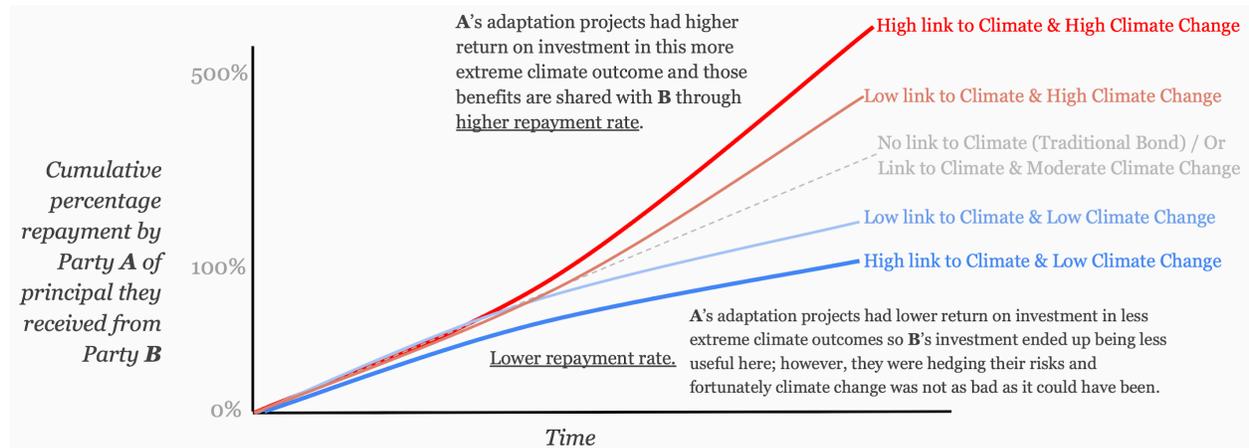

**Figure 9**: The cumulative percentage repayment of the principal for a traditional bond is not linked to the climate change that might occur during the life of the bond, rather, the amount repaid is purely a function of the interest rate – this is plotted as the grey dashed line. The extent to which repayment is linked to climate change and the amount of climate change both impact the amount repaid for climate contract bonds. If the climate ends up being more extreme, then the repayment schedule will be shaped like the one of the top two (red) lines. If the climate ends up being moderate, then the repayment schedule will be shaped like the one of the bottom two (blue) lines.

A climate-contingent repayment structure can be applied to any climate-related variable(s)[55] or combinations thereof at any time scale to fund any adaptation projects designed to reduce risk exposure to specific climate thresholds.[56] The potential scope of climate-contingent finance is vast. The next section focuses on sea-level rise in cities to illustrate the value that climate-contingent financing can provide.

---

[55] For instance, e.g., temperature, sea-level, precipitation or drought. Data from the U.S. government agencies, such as NOAA (e.g., https://data.noaa.gov/datasetsearch/), can be used as a trusted source for nearly any climate-related variable that would be useful to index to.

In selecting the climate variable, it is a balance between basis risk and broad applicability. Basis risk is the difference between the conditions under which a contract pays out and the conditions that one would like to hedge. For example, global sea-level rise levels may not translate perfectly into the benefits that a seawall provides compared to local relative sea-level rise, i.e., a contract tied to levels of global sea-level rise would have higher basis risk than a contract tied to local relative sea-level rise.

However, there is a trade-off with how widely applicable the climate variable is: the less potential entities that are impacted by the climate variable (e.g., local levels compared to global levels), the fewer opportunities there are to connect counterparties and facilitate hedging activity.

[56] For example: three inches of sea-level rise in 2030.

Forthcoming in *Berkeley Business Law Journal* 20## IV. CASE STUDY: CITIES & SEA-LEVEL RISE

Tens of trillions will likely be spent on infrastructure in the next ten years and it should be designed with climate change in mind.[57] There is scientific consensus that the sea-level has been rising and will continue to rise, but uncertainty remains regarding the future extent.[58] Coastal areas have experienced higher rates of relative sea-level rise than the global average.[59] Climate scientists expect a median 40-fold increase in severe flooding along the US coastline by 2050,[60] and there may be a 46% increase of global assets at risk of flooding due to extreme sea levels by 2100.[61]

Coastal areas are critical for the American economy and they face significant potential harm as a result of sea-level rise (SLR). More than half of the U.S. population lives or works in coastal counties, and these counties generate 58% of the U.S. GDP.[62] Up to 85% of local governments' revenue in the U.S. is generated from property taxes.[63] With (an extreme) 6.9 feet of SLR, 120 municipalities in the U.S. would risk losing 20% or more of their current property tax base, and 30 municipalities could lose as much as 50% of their property tax base.[64]

---

[57] Jonathan Woetzel et al., *Confronting Climate Risk*, MCKINSEY (2020), https://www.mckinsey.com/business-functions/sustainability/our-insights/confronting-climate-risk?cid=other-eml-alt-mcq-mck&hlkid=cc4c61660bde477599a5edacbd98aa7d&hctky=11761135&hdpid=022e9747-050f-4e29-841c-cd5abd666118.

[58] *See generally* MICHAEL OPPENHEIMER ET AL., IPCC SPECIAL REPORT ON THE OCEAN AND CRYOSPHERE IN A CHANGING CLIMATE, CHAPTER 4: SEA LEVEL RISE AND IMPLICATIONS FOR LOW-LYING ISLANDS, COASTS AND COMMUNITIES (2019), https://www.ipcc.ch/srocc/chapter/chapter-4-sea-level-rise-and-implications-for-low-lying-islands-coasts-and-communities/; *see generally New York City Panel on Climate Change 2019 Report: Chapter 3, Sea Level Rise*, 1439 N.Y. ACAD. SCI. 71 (2019), https://nyaspubs.onlinelibrary.wiley.com/doi/10.1111/nyas.14006.
And according to the 2019 IPCC review: "Comprehensive broad-scale projections of sea level at the coast including regional sea level changes, tides, waves, storm surges, interactions between these processes and accounting for changes in period and height of waves and frequency and intensity of storm surges are yet to be performed." *Id.* at 360. *See* Appendix B for more on sea-level rise.

[59] Jonathan Tirone, *Rising Sea Levels Inundating Coastal Economies Four Times Faster*, BLOOMBERG, Mar. 8, 2021, https://www.bloomberg.com/news/articles/2021-03-08/rising-sea-levels-inundating-coastal-economies-four-times-faster?sref=FUtuEW8l.

[60] Maya K Buchanan et al., *Amplification of Flood Frequencies with Local Sea Level Rise and Emerging Flood Regimes*, 12 ENV'T RSCH.: LETTERS 1, 1 (2017) https://iopscience.iop.org/article/10.1088/1748-9326/aa6cb3.

[61] *See generally* Ebru Kirezci et al., *Projections of Global-Scale Extreme Sea Levels and Resulting Episodic Coastal Flooding Over the 21st Century*, SCI. REPS., July 30, 2020, at 1, 6, https://www.nature.com/articles/s41598-020-67736-6.

[62] SUSANNE C. MOSER ET AL., U.S. GLOBAL CHANGE RESEARCH PROGRAM, COASTAL ZONE DEVELOPMENT AND ECOSYSTEMS, CLIMATE CHANGE IMPACTS IN THE UNITED STATES: THE THIRD NATIONAL CLIMATE ASSESSMENT, at 581 (2014), https://nca2014.globalchange.gov/report/regions/coasts. And, according to Alec Tyson & Brian Kennedy, 70% of Americans who live within 25 miles of the coastline say climate change is already affecting their community. Additionally, 57% who live 300 miles or more from the coastline say they have witnessed at least some impacts. Alec Tyson & Brian Kennedy, *Two-Thirds of Americans Think Government Should Do More on Climate*, PEW RSCH. CTR. (June 23, 2020), https://www.pewresearch.org/science/2020/06/23/two-thirds-of-americans-think-government-should-do-more-on-climate/.

[63] As a percentage of local own-source revenue (i.e., excluding transfers from state and federal governments), property taxes constitute 85% of local revenue inside Connecticut, for example. Linda Shi & Andrew M. Varuzzo, *Surging Seas, Rising Fiscal Stress: Exploring Municipal Fiscal Vulnerability to Climate Change*, 100 CITIES at 2 (May 2020), https://www.sciencedirect.com/science/article/pii/S0264275118314100.

Despite the importance of these areas, there is a shortage of funds available to take preventative measures to protect that value. A survey of 800 cities found that 43% of them did not have an adaptation plan.[65] Philip Stoddard, Mayor of South Miami, said, "Our first infrastructure challenge is going to be loss of septic tank function. . . . We are looking at the costs and cringing. Nobody is going to help, not the feds, not the state, not the county. So, cost is the biggest barrier."[66] In short, SLR is a massive problem for coastal cities,[67] but they lack financing options suited for the adaptation task.[68] Future tax revenue that otherwise would be much lower without adaptation under extreme SLR could provide a backing for municipalities to raise SLR-contingent financing as *A* parties.

### A. Vietnam (a Potential Adapter) and U.K. (a Potential Backer)

We draw on an analysis by McKinsey[69] that simulated future floods at high resolution in Ho Chi Minh City (HCMC), Vietnam and Bristol, UK.

HCMC is planning two hundred infrastructure projects for construction by 2050, including a metro system, power plants, water processing plants, port developments, and an airport. 45% of its land is less than one meter above sea level. Currently, there is low flooding risk to the existing

---

[64] *Underwater: Rising Seas, Chronic Floods, and the Implications for US Coastal Real Estate*, UNION OF CONCERNED SCIENTISTS 1, 5 (June 18, 2018), https://www.ucsusa.org/resources/underwater.

[65] Fiona Harvey, *One in Four Cities Cannot Afford Climate Crisis Protection Measures – Study*, THE GUARDIAN, (May 21, 2021), https://www.theguardian.com/environment/2021/may/12/one-in-four-cities-cannot-afford-climate-crisis-protection-measures-study.

[66] UNION OF CONCERNED SCIENTISTS, *supra* note 64, at 19.

[67] "On the [municipal] revenue side, SLR could fundamentally restructure local economies and erode property taxes. [...] On the expenditure side, added local costs include: maintaining and repairing roads due to coastal storm events and rising water tables; adapting water supply and drainage systems to account for more intense storms and storm surge; and expanding community health, education, and disaster preparedness and response." Linda Shi & Andrew M. Varuzzo, *Surging Seas, Rising Fiscal Stress: Exploring Municipal Fiscal Vulnerability to Climate Change*, 100 CITIES 1, 2 (May 2020), https://www.sciencedirect.com/science/article/pii/S0264275118314100.

[68] For more on this topic, see generally Marcus Painter, *An Inconvenient Cost: The Effects of Climate Change on Municipal Bonds*, 135 J. FIN. ECON 468 (2020), https://www.sciencedirect.com/science/article/abs/pii/S0304405X19301631 ("Counties more likely to be affected by climate change pay more in underwriting fees and initial yields to issue long-term municipal bonds compared to counties unlikely to be affected by climate change").

Furthermore, regulations that would enforce environmental management related to climate adaptation have also been documented to raise the cost of capital for municipal bond issuers. *See* Akshaya Jha et al., *Polluting Public Funds: The Effect of Environmental Regulation on Municipal Bonds* 45 (Nat'l Bureau Econ. Rsch. Working Paper, Paper No. 28210, 2020), https://www.nber.org/papers/w28210.

[69] *Can Coastal Cities Turn the Tide on Rising Flood Risk?*, MCKINSEY GLOB. INST. (Apr. 20, 2020), https://www.mckinsey.com/business-functions/sustainability/our-insights/can-coastal-cities-turn-the-tide-on-rising-flood-risk.

infrastructure but a flood with the same probability in 2050 would have three times the physical infrastructure damage and twenty times the knock-on economic effects.[70]

Billions are earmarked for spending on elevating certain high-risk metro stations. Let us say the elevation projects are being designed to perform in conditions of up to 0.5 meters of sea-level rise. There is a plausible scenario, though, that sea level will be one meter higher within the life of the metro stations. The one-meter scenario is not deemed a high probability in the near-term and will therefore be difficult to justify raising debt to finance additional protection to guard against.

In Bristol, by 2065, a 0.5% probability flood would produce eighteen times more infrastructure damage and thirty times more knock-on effects compared to a 0.5% probability flood today. However, "unlike in Ho Chi Minh City, most of the infrastructure the city plans to have in place in 2065 has already been built. In the short term, Bristol's hands are likely largely tied . . . The city is still scoping out a range of options to protect the city."[71]

It is usually more cost effective to build new infrastructure that meets higher standards than to retrofit existing infrastructure.[72] Infrastructure projects in the planning phase can be built to defend against more extreme climate change through upfront spending. Bloch et al. (2012) point out that, in doing so, the owner is implicitly buying out-of-the-money options on more extreme climate change, allowing the city to, effectively, generate future cash flows under extreme climate change. The owner can then monetize those future cash flows by entering into a climate contract for part of the financing of the project.[73] Since retrofitting is more expensive, it may make more sense for existing infrastructure owners to delay a decision to adapt, and instead take the other side of this trade by buying protection to finance a future retrofit or rebuild.[74]

HCMC could build to a one-meter SLR specification to provide protection against a more extreme than expected outcome by issuing traditional debt to pay for the first 0.5 meters of protection and selling a climate contract to pay for the additional 0.5 meters. Bristol, with much of its infrastructure already in place, could buy part of that climate contract. Other long-term capital with similar exposures, like pension funds and owners of real assets, similarly could benefit from buying part of such climate contracts. If SLR is more than one meter before 2030, HCMC pays investors a predetermined return on their investment, which the investors could use to finance part of their infrastructure retrofitting (or supplement their returns).

This simple illustrative example demonstrates clear benefits for HCMC, reducing their risk physically, and Bristol, reducing their risk financially. However, it is unlikely that cities would serve as direct counterparties to one another in this manner. We discuss the more likely sets of counterparties in Section V.

---

[70] In 2050, a one percent probability flood in a 180-centimeter relative sea-level-rise simulation would lead to widespread infrastructure damage with 66% of the city underwater. *Can Coastal Cities Turn the Tide on Rising Flood Risk?*, MCKINSEY GLOB. INST. 13 (Apr. 20, 2020), https://www.mckinsey.com/business-functions/sustainability/our-insights/can-coastal-cities-turn-the-tide-on-rising-flood-risk.
[71] *Id.*
[72] *See generally* Hallegatte, *supra* note 4, at 241.
[73] *See* Bloch et al. (2012), *supra* note 36, at 89-90, 96-97.
[74] *See id.* at 89-90, 95.

*B. New York*

Over the last 10 years, every county in the state of New York has been impacted by flooding or extreme storms.[75] The New York City metropolitan area is exploring its options for infrastructure to reduce coastal flooding and storm risks due to climate change. The U.S. Army Corps of Engineers (USACE) calculated relative sea-level change (RSLC) projections for the area.

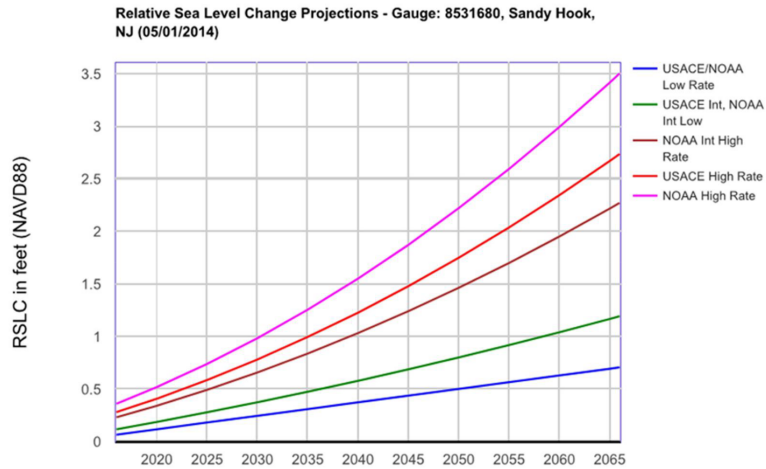

**Figure 10:** RSLC projections.[76]

The USACE notes that:

> Benefits were developed on the intermediate rate of RSLC. It has been observed that use of the low or historic rate of RSLC will favor perimeter measures in plan selection, while the use of the high rate of sea level change favors larger barriers . . . the next round of formulation will also include a detailed investigation how the alternatives perform under each of the RSLC scenarios.[77]

They computed net benefits of $124 billion for the proposed infrastructure project. The cost-benefit analysis is dependent on the assumption of an intermediate rate of RSLC. However, further adaptation projects place them in the position of an *A* effectively buying out of the money options, as contemplated above. To fund the parts of the project that build additional height to the coastal defenses required for that level of SLR, they could engage in a financing that is indexed to

---

[75] *Digital Dialogue No. 5: Scaling Up Coastal Ecosystem Protection*, WHARTON RISK MGMT. AND DECISION PROCESSES CTR. (July 2020), https://riskcenter.wharton.upenn.edu/digital-dialogues/coastalecosystemprotection/.

[76] U.S. ARMY CORPS OF ENG'RS, N.Y. DIST., N.Y.-N.J. HARBOR & TRIBUTARIES COASTAL STORM RISK MANAGEMENT INTERIM REP. 111 (Feb. 2019), https://www.nan.usace.army.mil/Portals/37/docs/civilworks/projects/ny/coast/NYNJHAT/NYNJHAT%20Interim%20Report%20-%20Main%20Report%20Feb%202019.pdf?ver=2019-02-19-165223-023.

[77] *Id.*

Forthcoming in *Berkeley Business Law Journal*                                          24the rate of SLR underpinning their cost-benefit analysis. We explore who their potential ***Bs*** might be below.

## V. CONTRACT PARTICIPANTS

Although cities like Bristol stand to benefit from becoming ***B*** parties to these instruments, a much wider array of actors would also benefit from becoming ***Bs*** for ***As***. Regional banks that own mortgages on at-risk properties,[78] a variety of public and private long-term pools of capital,[79] institutional real estate investors with high exposure to sea-level rise and other climate change induced hazards;[80] and federal and state governments all stand to significant benefit from these instruments. ***Bs*** include entities that own similar real assets as ***As*** but are not in a position to make physical adaptations in the near-term (e.g., Bristol in the UK, or long-term real estate investors), and entities that would end up serving as a financial backstop to ***As*** (e.g., governments and insurance companies).[81]

***As*** are entities looking to make physical changes now, usually for one of two reasons. First, some ***As*** would like to physically reduce their climate risk, like by building new infrastructure that resists extreme weather conditions. Second, other ***As*** stand to benefit from climate change and are

---

[78] NY Department of Financial Services: "Regional and community banks . . . are more vulnerable to regionally concentrated physical risk, including to sudden extreme events . . . These banks' property loans tend to be more geographically concentrated than the loans of larger banks. In addition, CRE [commercial real estate] loans constitute a much larger share — nearly a third — of the loan books of small banks." Dan Ennis, *NY Regulator Lays Out Climate Risk Expectations for Banks*, BANKINGDIVE (Oct. 30, 2020), https://www.bankingdive.com/news/Department-Financial-Services-new-york-climate-risk-expectations-stress-test/588103/.

Jonathan Woetzel et al., *Will Mortgages and Markets Stay Afloat in Florida?*, MCKINSEY GLOB. INST. (Apr. 27, 2020), https://www.mckinsey.com/business-functions/sustainability/our-insights/will-mortgages-and-markets-stay-afloat-in-florida ("local and regional banks that own concentrated portfolios of mortgages on coastal properties may find themselves especially vulnerable to near-term climate events"). Freddie Mac finds that SLR could "destroy billions of dollars in property and displace millions of people," with impacts greater than "the housing crisis and Great Recession." UNION OF CONCERNED SCIENTISTS, *supra* note 64, at 15. Paulo Issler finds that wildfires and flooding cause increased residential mortgage default rates. Paulo Issler et al., Mortgage Markets with Climate-Change Risk: Evidence from Wildfires in California (July 1, 2020) (unpublished manuscript), https://papers.ssrn.com/sol3/papers.cfm?abstract_id=3511843.

[79] Including, for example, pension funds, endowments, sovereign wealth funds, and private investment firms that own coastal real assets.

Divya Mankikar, an investment manager at the California Public Employees' Retirement System, the country's biggest public pension fund, said in September 2020: "Climate change is one of the top three risks to our fund. We pay pension and health benefits to over two million current and former state employees. So, the payout is decades out." Coral Davenport & Jeanna Smialek, *Federal Report Warns of Financial Havoc from Climate Change*, N.Y. TIMES (Sept. 8, 2020), https://www-nytimes-com.cdn.ampproject.org/c/s/www.nytimes.com/2020/09/08/climate/climate-change-financial-markets.amp.html.

[80] At least $130 billion of U.S. institutional real estate is located in the U.S. coastal areas that are in the top 10% for exposure to sea-level rise, according to *Futureproofing Real Estate from Climate Risks: New Research from ULI in Partnership with Heitman*, HEITMAN (Oct. 9, 2018), https://www.heitman.com/news/futureproofing-real-estate-from-climate-risk-new-research-from-uli-in-partnership-with-heitman/.

[81] The third type of ***B***, which is relevant if/when these contracts are widely marketed, are entities that believe they have superior climate projection information and would expect to better estimate the expected value of a contract, what the financial economics literature calls "speculators."

making changes in expectation. For example, owners of land in areas that are currently too cold for growing certain crops may expect to start to be in the correct climate zones as the climate changes and are making changes to take advantage of that possibility, should it arise.[82]

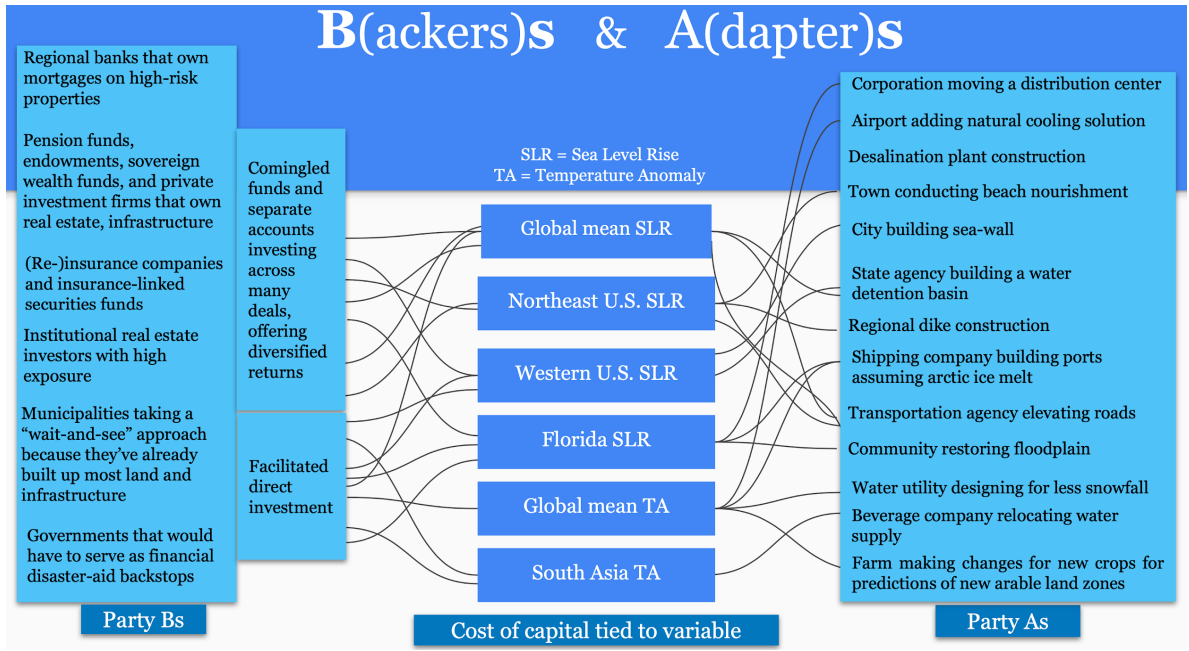

**Figure 11:** Conceptual diagram of Climate Contracts linked to thresholds on sea-level rise (SLR) and temperature anomaly (TA) values. Examples of Limited Partners in a set of commingled funds or separately managed accounts that invest in **B** positions in the climate contracts and examples of parties to the **A** side of the contracts are listed. In addition to *A* and *B* market participants hedging exposures, there will likely also be investors participating in order to achieve returns uncorrelated from traditional asset classes and potentially actors stabilizing markets[83]. Many **B** participants will view their participation in the fund(s) as both a hedge and an opportunity for an uncorrelated real return. There would be essentially no capacity constraint on the size of the fund(s); they could invest hundreds of billions of dollars without decreasing expected returns.

Long-term investors like sovereign wealth funds, public pensions, family offices, insurance companies, and endowments seek diversifying investment returns and would be attracted to returns from a fund that invested as a ***B*** party in climate contracts.[84] Institutional investors search for return streams where they can model the correlation to other return streams, or at least qualitatively understand the potential correlations. The climate system, although full of uncertainty, is more amenable this scientific analysis than the economic and financial systems driving the price of many alternative investments: "After 2008, ... it had become clear that risk analyses based on the physics of the earth and atmosphere were far more reliable than the assumptions around human herd behavior implicit in the risk of subprime mortgage markets or Russian bond defaults."[85] Climate trends are also slower moving than most other financial variables; however, the trends are still fast enough that they occur within the lifetime of an infrastructure project and therefore need to be accounted for in the financing of that infrastructure.

---

[82] "[T]he introduction of heat-tolerant varieties of grapes, could sharply reduce the losses [in the wine production industry] in California and turn the Washington loss into a 150% gain." Richard A. Kerr, *Vital Details of Global Warming Are Eluding Forecasters*, SCIENCE, Oct. 14, 2011, at 173-74, https://www.science.org/doi/abs/10.1126/science.334.6053.173; *see also* J. B. Ruhl, *The Political Economy of Climate Change Winners*, 97 MINN. L. REV. 206, 217-31 (2012).

[83] *See* Reuters Staff, *In First for Fed, U.S. Central Bank Says Climate Poses Stability Risks*, REUTERS (Nov. 9, 2020, 1:03 PM), https://www.reuters.com/article/us-usa-fed-stability-climate-idUSKBN27P2T9 (quoting U.S. Federal Reserve Governor Lael Brainard statement from November 2020: "Chronic hazards, such as slow increases in mean temperatures or sea levels, or a gradual change in investor sentiment about those risks, introduce the possibility of abrupt tipping points or significant swings in sentiment.").

Resulting mark-to-market pricing data could guide policy-makers on market expectations of climate outcomes. If there is eventually secondary trading of underlying contracts, prices on specific contracts[86] would reveal up-to-date estimates of specific climate risks, globally guiding public policy and planning.[87] In a simulation model, market participation caused traders to converge quickly toward believing the "true" climate model, suggesting that a market for climate contingent instruments could be useful for building public consensus on climate solutions.[88] Schlenker and Taylor (2019) found that weather derivatives have been pricing in warming trends approximately in line with climate model projections, which suggests that markets are pricing in some climate change information and that information can be extracted from market price data.[89]

## VI. SIMULATIONS OF CLIMATE CONTRACTS

To begin, we explain the mechanics of climate contracts, and incrementally build complexity to demonstrate possible outcomes. Climate contracts are a type of derivative instrument where the underlying financial asset is a climate index. We investigate climate contracts through simulations where there are two types of counterparties, ***As*** and ***Bs***. For example, in a simple CC a ***B*** provides an ***A*** with capital that it can use from outset. In return, the ***A*** owes an obligation to pay back the principal and a one-time payout to the **B**, which we call the "price" of the contract, if the climate scenario or one more extreme is realized before expiration of the CC.

---

[84] Family offices are increasingly focused on climate change. "This is not only because of the views of younger heirs — millennials and Gen-Z members are frequently characterized as more environmentally and socially aware than their elders. It is also down to a heightened awareness of the impact of climate change risks on long-term investment portfolios — something family offices, with their objectives of preserving wealth for future generations, are particularly concerned about." Alice Ross, *Climate Concerns Reaching 'Tipping Point' for Family Offices*, FIN. TIMES (Oct. 13, 2020), https://www.ft.com/content/692a8f67-325a-4e93-ac8a-6717c011d0b3. "More net money flowed into ESG funds between April and July 2020 than the *entire* previous five years." Michael J. Coren, *BlackRock is Forcing Finance to Take Climate Risk Seriously*, QUARTZ (June 10, 2021), https://qz.com/1957979/blackrock-is-forcing-wall-street-to-take-climate-risk-seriously/.
[85] ROBERT MUIR-WOOD, THE CURE FOR CATASTROPHE 144 (Basic Books 2016).
[86] E.g., SLR at 2 inches in 2030.
[87] *See generally* John J. Nay et al., *Betting and Belief: Prediction Markets and Attribution of Climate Change*, 2016 WINTER SIMULATION CONFERENCE 1666 (2016), https://ieeexplore.ieee.org/abstract/document/7822215.
[88] *See generally id.* The simulation model exploring traders' beliefs about the cause of climate change can be found at https://github.com/JohnNay/predMarket.
[89] Wolfram Schlenker & Charles A Taylor, *Market Expectations About Climate Change* 18-19 (Nat'l Bureau of Econ. Rsch. Working Paper 25554, 2019), https://www.nber.org/papers/w25554.

Terms are specified when the CC is initially sold, including the initial amount that the **B** pays to the **A**,[90] the extreme climate scenario which triggers the one-time payout,[91] the amount that the **A** would pay the **B** if the defined climate scenario is realized,[92] and the term within which the climate threshold must be passed to cause the payout.[93]

*A. Simple Setup*

In this simple setup, **A** and **B** only issue and purchase contracts for the most extreme climate scenario, which has a probability of *p*, realized (or not) on a time horizon of *y* years. Focusing only on the extreme climate scenario first allows us to demonstrate the basics of the pricing relationships.[94] **A** finances adaptation by selling a CC. **B** has two methods to financially hedge climate impacts: invest in risk-free assets that return *s* per year, or buy a CC. At the beginning of each time period, they engage in a CC at an amount of their initial assets (e.g., $1e8). At the end of the specified time period, two outcomes are possible. If the extreme climate scenario occurs, then the assets **A** invests in adaptation provide a return of *x*. If the CC is not triggered then the contract expires and the return is zero.[95]

We set parameters to the following values, *s* = 1%, *p* = 0.1, *y* = 10 and vary the values for the CC price and *x*. The conclusions are not dependent on the specific values of these variables; they are just for illustration. We then simulate ten ten-year time periods (i.e., 100 years). At the end of each period, we simulate the notion that definitions of climate scenarios (e.g. "extreme," "intermediate") are adjusted to reflect NOAA's definitions as of that time. Given these parameters, over each ten-year time period **B** would earn a 10.46% risk-free return. The risk-free return (10.46%) divided by the probability of the scenario **B** would like to hedge (0.1) is the *minimum* price for a contract (11.046), because if the price was any lower **B** would invest in risk-free assets for a higher expected dollar outcome.

In Figure 12, we plot a ten-time-period simulation with zero, one, and two realizations of the extreme scenario. Each ten-year time period is independent from the other nine. The run with one extreme scenario realization (middle chart) is reflective of the expected value of assets for **B** under these settings. The price is set to the minimum acceptable CC price for **B** (11.046), so the assets of **B** are equivalent to their assets if they had invested risk-free. If the extreme scenario is realized less than expected (left chart), **B** is worse off investing in CCs than investing risk-free. If the scenario is realized more than expected (right chart), **B** is better off investing in CCs than investing risk-free.

---

[90] E.g., $1e8 ($100 million).
[91] E.g., an "extreme scenario" as defined by NOAA at the time of issuance.
[92] E.g., 7x $1e8.
[93] E.g., 10 years.
[94] We will specify the scenarios for actual CCs. For this simulation, we do not discuss the specifics of defining the scenarios.
[95] We discuss below the realistic assumption that adaptation investments also provide some return in less extreme scenarios, which makes investing in adaptation more attractive for *As*.

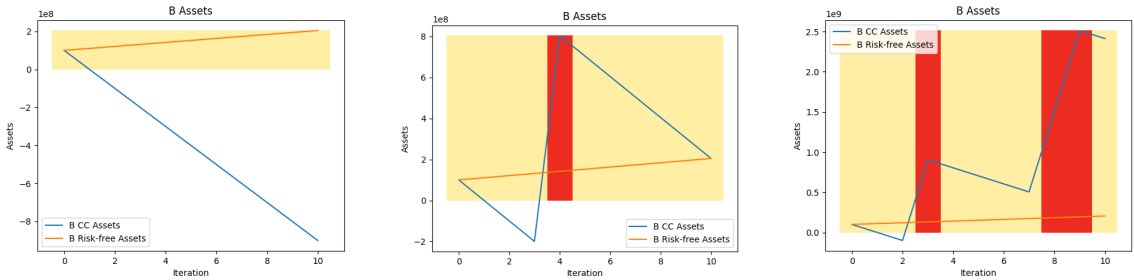

**Figure 12:** **B**'s assets under varying number of times of extreme climate scenario realizations; zero (left), one (middle), and two (right). The price for a CC on the extreme scenario was set to 11.046.

**B** would likely engage in the CC transactions where the utility they derive from wealth is a positive function of the extremity of the climate outcome during the period they obtain the wealth. In other words, **B** invests when the expected final wealth of CCs are designed to be equal to the risk-free investment option because the large monetary payoff from CCs in the extreme climate scenario is worth more than a payoff of equal dollar value in situations where the climate is better.[96] Furthermore, before the triggering of the contract, **B** can operate with less financial climate risk, and therefore more creditworthiness, with these contracts in place.

In Figure 13, we plot a simulation with the price of the CC at 15 and one extreme scenario occurs. If the price is above 11.046, **B**'s expected asset values are greater when investing in CCs than investing risk-free. If higher CC prices are acceptable to **A**, then **B** can be made much better off relative to risk-free investing. Next, we turn to CCs from **A**'s perspective to see if this is the case.

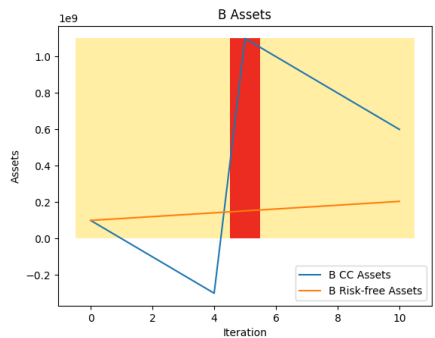

**Figure 13:** **B**'s assets in a ten-time-period-simulation where the price for an extreme scenario contract is 15 and the extreme scenario was realized once, which is the expected value number of times given the settings for the climate probability distribution.

---

[96] *See* Giglio et al., *supra* note 50, at 30.

For *A*, the price of the contract and the benefits of adaptation determine whether selling a CC is of positive expected value. Let us assume *A* obtains a 7x return on adaptation investments in the extreme scenario.[97] We set the price of the contract to 11.046, the lowest price a counterparty would accept.

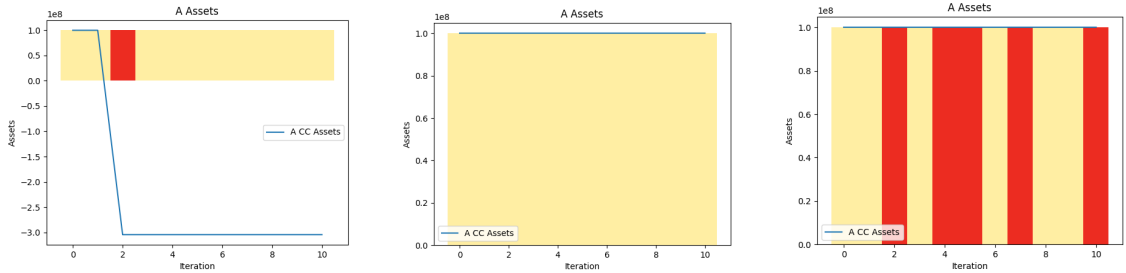

**Figure 14:** *A*'s assets with price for extreme scenario contract at 11.046. Return on adaptation investments set at 7 (left chart),11.046 (right chart), and the middle chart shows *A*'s assets regardless of the contract price or benefits payoff function: if the extreme scenario is not realized, *A*'s assets do not change.

If the contract price is greater than *A*'s return on adaptation investments (Figure 14, left chart), then *A* has less wealth after a contract is paid out. If the adaptation returns are modeled as equal to the price, *A* has the same wealth, regardless of whether a contract was paid out (Figure 14, right chart). If the extreme scenario is not realized, *A*'s assets do not change, regardless of the contract price or benefits payoff function because no contract is triggered (Figure 14, middle chart).

Thus far, we have modeled two assumptions that make CCs an unrealistic, unattractive choice for *A*; we are only modeling two climate scenarios (extreme and not extreme), and we are not modeling any benefits that adaptation investments provide beyond the end of a CC term.

## B. Additional Climate Scenarios

When buying a CC that pays out in an extreme scenario, subsequently realizing another scenario would still provide some benefit to *A* from the proceeds of selling that CC. To demonstrate these incremental returns, we model four scenarios in between the most and least extreme: *high, intermediate (int) high, int, and int low.*

| Climate Scenario | Probability of Scenario | Cumulative Probability | Contract Price | Capital Allocation 1 | Capital Allocation 2 | Adaptation Return |
|---|---|---|---|---|---|---|
| *low* | 0.3 | 1.0 | 1.105 | 0 | 0 | 0 |

---

[97] "[T]here is large potential for coastal adaptation to reduce the expected impacts of SLR compared to the alternative of no adaptation, lowering global net present costs through 2100 by a factor of seven . . . ." Diaz, *supra* note 39, at 1. "Each dollar of extra preparedness spending reduced disaster impacts by an average of $7 over a single four-year election cycle and disaster costs overall by an average of $15." MUIR-WOOD, *supra* note 85, at 159.

| | | | | | | |
|---|---|---|---|---|---|---|
| *int low* | 0.2 | 0.7 | 1.578 | 0 | 0.2 | 1.5 |
| *int* | 0.1 | 0.5 | 2.209 | 0 | 0.2 | 2.25 |
| *int high* | 0.1 | 0.3 | 3.682 | 0 | 0.2 | 3.75 |
| *high* | 0.1 | 0.2 | 5.523 | 0 | 0.2 | 5.5 |
| *extreme* | 0.1 | 0.1 | 11.046 | 1.0 | 0.2 | 7.0 |

**Table 1**: Simulation settings. Probabilities are the likelihood of realizing the scenario during each time period. Cumulative probabilities are the likelihood of triggering the scenario (being at or above that scenario). Prices are computed by dividing the risk-free return that is available to *B* for the ten-year time period (10.46%) by the cumulative probability of the scenario (and rounding to three decimal places).

We model adaptation benefits according to the "Adaptation Return" column in Table 1, assuming the less extreme the scenario the lower the return on a dollar invested in adaptation.

In the simulations above, we set the probability of the climate scenario for "*low*" to be 0.9 and "*extreme*" to be 0.1. Then, we model the probability of each scenario occurring for any iteration of a simulation according to the "Probability of Scenario," which assumes more extreme scenarios are rarer. We continue to have both parties engage in a CC at the beginning of each time period indexed to the extreme scenario at an amount of their initial assets ($1e8) and the minimum acceptable price for a risk-free investment by *B* (11.046). The only difference here (in Figure 15) from the simulation figures above is that we added additional scenarios indexed to the extremity of the climate scenario.

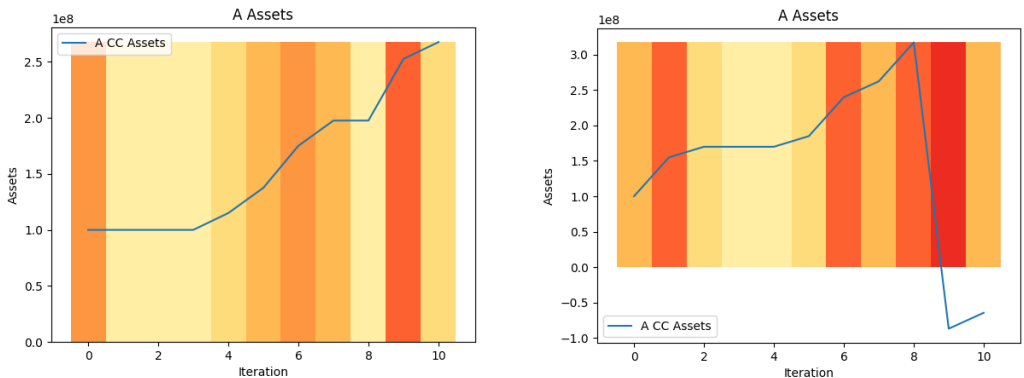

**Figure 15:** *A*'s assets with additional scenarios where adaptation investments can pay off. Darkness of the background color corresponds to the extremity of the climate scenario that iteration. The extreme scenario (dark red) did not occur in the simulation on the left and it occurred once in the simulation on the right.

Each chart in Figure 15 is a single ten-period simulation, which is helpful for explanatory purposes. However, to help understand the expected value for *A* over many simulations, we run a large batch and investigate the distribution of end-state outcomes. In the following analyses, we run 500 replications of a simulation of five ten-year time periods with the model parameters fixed. The only stochasticity is in the climate scenario realized (according to the Table 1, "Probability of Scenario" column) at each time period of each simulation.

We estimate from the mean of the simulated outcomes (dashed line in Figure 16), a positive expected value of the change in *A*'s assets from initialization of a simulation to the end. This positive expected value is due to the returns to adaptation that *A* enjoys in the non-extreme scenarios where *A* is not paying out any contracts. Money raised from a CC indexed to a higher-scenario contract is not paid back when a lower scenario is realized, because the higher scenario was not triggered.

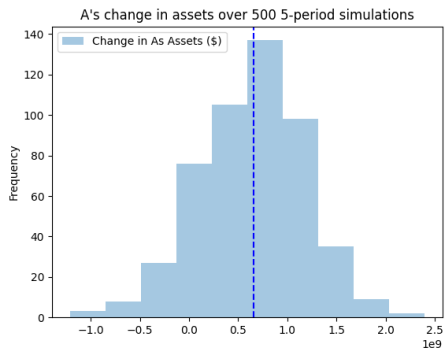

**Figure 16:** Distribution of *A*'s change in assets when only selling contracts for the *extreme* scenario over 500 runs of five-time-period simulations.

*A* can sell contracts for other scenarios. Let us take an example from the other end of the spectrum: *A* sells contracts for just the *int low* scenario each time period. They are now paying out the contract every time they receive a benefit from investing the proceeds because we are only modeling benefits from adaptation in scenarios equal to or greater than the extremity of the scenario they sold a contract for. However, there is approximately the same positive expected value for *A* (Figure 17) as there was in Figure 16. The positive expected value is due to the returns to adaptation realized in more extreme scenarios being higher than the price they pay for the *int low* contract payout. We have modeled money raised from a contract that is triggered in a lower scenario as being used for computing adaptation returns in a higher scenario, but it triggers a payment from *A* to *B* at the contract price (which is indexed to the lower scenario and therefore less expensive).

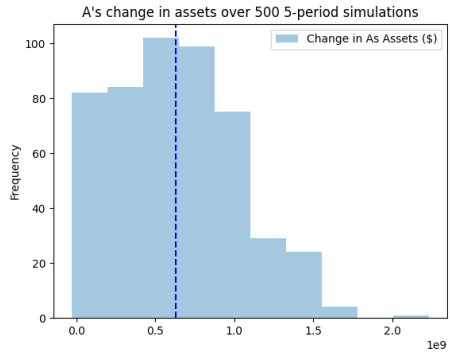

**Figure 17:** Distribution of *A*'s change in assets with additional scenarios when only selling contracts for the *int low* scenario, across 500 runs of five-time-period simulations.

*C. Non-fungible Adaptation Value*

Calculations for returns on adaptation are based on the climate scenario realized during the observed period. We are modeling dollars invested from the proceeds of CCs as fungible with respect to their returns on investment in climate scenarios for which they were not explicitly earmarked. In other words, dollars from climate contracts in more extreme (or less extreme) climate scenarios are invested and realize their returns according to the lower (or higher) return in the less extreme (or more extreme) climate scenario. It is unlikely that the assets raised for a given scenario would all be deployed in a manner that provides the same return when the eventual realized scenario is a different one.

To capture this reality, we model a reduction of value in the assets raised for a more extreme scenario than what is realized. If, in Time 1, *A* raised $10 by selling a contract that would trigger payout to *B* in an extreme scenario and in Time 2, the climate scenario ended up being intermediate (midpoint between the least and most extreme scenarios), we multiply the $10 by the appropriate discount factor (0.5), to model $5 of adaptation investment being applied to the calculation of return on investment (ROI) in that intermediate scenario. If the realized climate scenario is less extreme than the scenario the money was raised for, then there is a possibility that *A* inefficiently overbuilt adaptation projects with that capital, making the capital less effective at delivering benefits in a milder climate world. This is the "*upper scenario discount*" we use to model overpreparation.

If the realized scenario is more extreme than the scenario the money was raised for, we can also model a reduction of value in the capital invested from the proceeds of selling a CC indexed to that milder scenario since *A* may not have prepared efficiently for that more extreme scenario. We call this the "*lower scenario discount*."

In the next set of simulations, we have *A* and *B* allocate equal amounts (20%) to the five scenarios above the *low* scenario. In the simulations above, *A* and *B* both allocated 100% to only one scenario. The top left chart in Figure 18 displays the outcome across varying upper scenario discounts and the top right chart displays the outcome across lower scenario discounts.

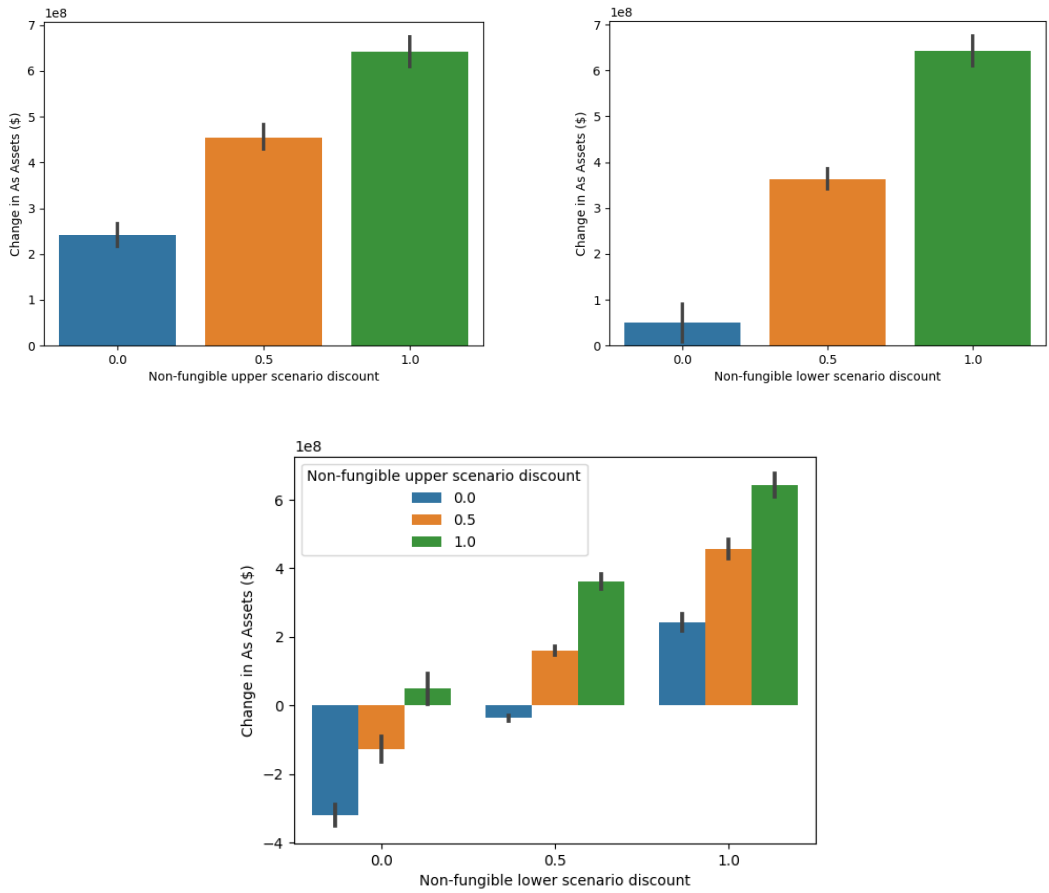

**Figure 18:** *A*'s change in assets while varying the upper scenario discount and keeping the lower scenario discount fixed at a value of one (upper left chart) and while varying the lower scenario discount and keeping the upper scenario discount fixed at one (upper right). The bottom chart displays *A*'s change in assets under all nine combinations of parameter values. The small lines in these charts are confidence intervals computed by 1,000 bootstrapped samples of the simulation outcome data.

The adaptation investments would not be useless in scenarios that they were not explicitly designed for. In addition, some value may be realized early in the timeline (e.g., credit rating improvement). Therefore, a value of zero is unrealistically low for both types of discounts. However, investments designed for a more extreme scenario than what occurs is likely more problematic for the full realization of its benefits compared to investments designed for a scenario less extreme than what occurs. Take the example of a sea-wall. If it is built to withstand fifteen feet of storm surge, but only ten feet occurs, then it was overbuilt and the capital used to build that additional protection is essentially wasted; therefore, the upper scenario discount factor is less than one. Contrast that to building to withstand ten feet and storm surge actually being fifteen feet. In this case, building for ten feet likely has nearly 100% of the capital used to build it providing a return on its investment because realized damages are significantly lower than damages in the same scenario without a sea wall; therefore, the lower scenario discount factor is high, probably closer to one. This suggests that, although both are greater than zero, the *lower scenario discount factor* we use in the simulations should likely be higher than the *upper scenario discount factor*. Given this understanding, we set the upper scenario discount to 0.5 and the lower scenario discount to 0.75 for the subsequent analyses.

### D. Historical Adaptation Value

Next, we add a realistic assumption that *A*'s adaptation investments provide benefits beyond the end of the contract time-frame (in the cases modeled thus far, ten years). To model degradation of the adaptation projects, we multiply the return on adaptation benefits by a discount factor determined by when the investments were made. The left chart in Figure 19 visualizes the function used to map time elapsed since a project's implementation (in years) to the discount factor multiplied by the adaptation ROI that would be realized from the dollars invested in those prior time periods. We start at a factor of almost one for the present and end at a factor near zero for four time periods (40 years) ago.

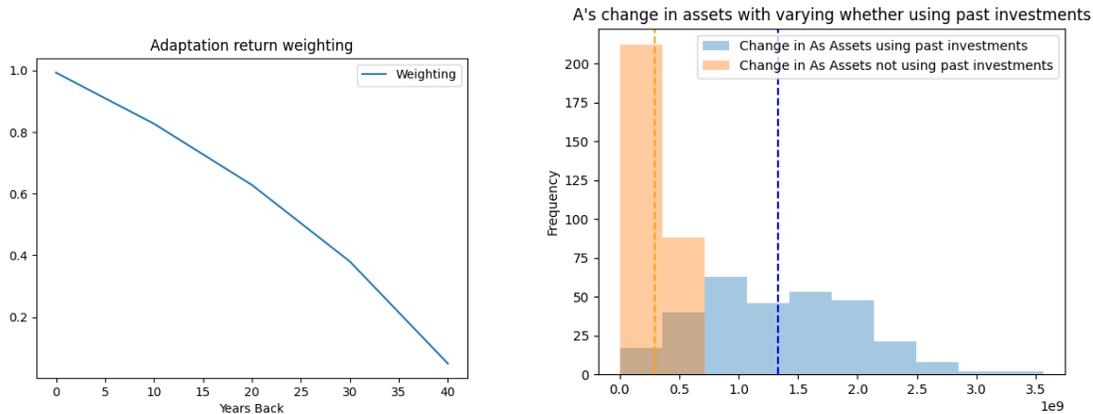

**Figure 19:** Adaptation return factor as a function of time (left chart), and *A*'s change in assets varying whether past investments are modeled as having value (right chart). In this set of simulations, *A* and *B* allocate equal amounts (20%) to the five scenarios above *low*; lower scenario discount is 0.75 and upper scenario discount is 0.5; further simulation settings can be found in Table 1.

Modeling past investments as if they have value in subsequent time periods has a large, positive impact on *A*'s change in expected assets over the course of a five-period (fifty-year) simulation.

### E. Outcome Measures

Before analyzing CC pricing, we discuss the simulation outcomes we are measuring and optimizing for in the next section. *A* is in a position to make physical changes today that could help avoid ruin in unlikely scenarios, but cannot justify the capital expenditure in situations where those scenarios do not occur. *A* will pay a high price for the avoidance of ruin, but only when it is proven useful to do so.

Implemented climate-contingent instruments may be a hybrid between pure debt and pure CCs, but for explanatory purposes in this Section we discuss pure debt and pure CCs to illustrate the distinction. If the climate effects turn out better than anticipated, *A* incurs a lower cost of adaptation capital from CCs than from debt. If the climate effects turn out worse than anticipated, debt could have a lower cost of capital than a CC product. However, *A* is explicitly hedging against a scenario that they believe is unlikely; otherwise, they would have raised all the capital through traditional debt. We are specifically analyzing situations where *A* could not have raised pure debt, either because their internal stakeholders (e.g., city council members) or potential debt investors would not be interested.[98] By offering CCs, *A* is only paying for what they need.

*A* will likely issue pure debt for projects with more assured benefits that accrue across most of the plausible climate change scenarios, rather than sell a CC. But the express purpose of CCs is to finance projects that account for various climate scenarios, some which may have a relatively high likelihood of non-occurrence — these are not the types of projects that are usually politically or economically feasible to raise pure debt for. Entities that do not issue their own sovereign currencies (e.g., cities, private sector entities such as corporations, U.S. states, and countries in the E.U.) cannot issue unlimited debt.[99] Municipal and state governments often have limits on the amount of debt they can issue, and companies have equity shareholders that can be disadvantaged by additional debt in downside scenarios. *A*s are motivated to utilize CCs in part because they cannot pay for protection from the low probability climate scenarios. Given that most scenarios they are protecting against may not occur, it is difficult to justify spending for such protection. If the spending can be tied to whether the scenario occurs, it is more palatable to stakeholders involved in the capital planning. Therefore, the most appropriate outcome to initially measure for *A* is the difference between their assets at initialization of a simulation and their assets at the end of the simulation.[100]

For *B*s, we measure the difference between their assets at the end of the simulation if they had invested in CCs and their assets at the end of the simulation if they had made risk-free capital allocations. *B*s have exposure to the same climate scenarios, but do not have the capacity or willingness to make physical changes that would avert the potential damages. Given these factors and the possibility of investing risk-free from Time 1, risk-free investing is the relevant opportunity cost to benchmark their investment outcomes.

---

[98] In addition to the scientific uncertainty that exists around climate change, there are a large number of people in the U.S. that do not believe any significant human-driven climate change will occur. Such individuals would likely be thrilled for their company or city to raise capital for which the repayment is contingent on climate change occurring.

[99] The only entities that issue and borrow in their own fiat sovereign currencies are the national-level governments in the U.S., U.K., Canada, Japan, Australia, and a few other countries.

[100] Rather than, for example, the difference between the actual outcome with the CC and a hypothetical outcome with a pure debt raise because, as explained in this paragraph, it is unlikely they would have raised traditional debt for this.

*F. Price Optimization*

We set the probability of each scenario's occurrence and *A*'s return on adaptation investments in each scenario to the same values as before (see Table 1): *A*'s upper scenario discount to 0.5, lower scenario discount to 0.75, and historical discount to the same levels that were employed in the simulations illustrated in Figure 14. We also set *B*'s risk-free rate of return to the same value, 1%. Figure 20 visualizes *A* and *B*'s respective outcomes for these settings under the contract prices we have used thus far.

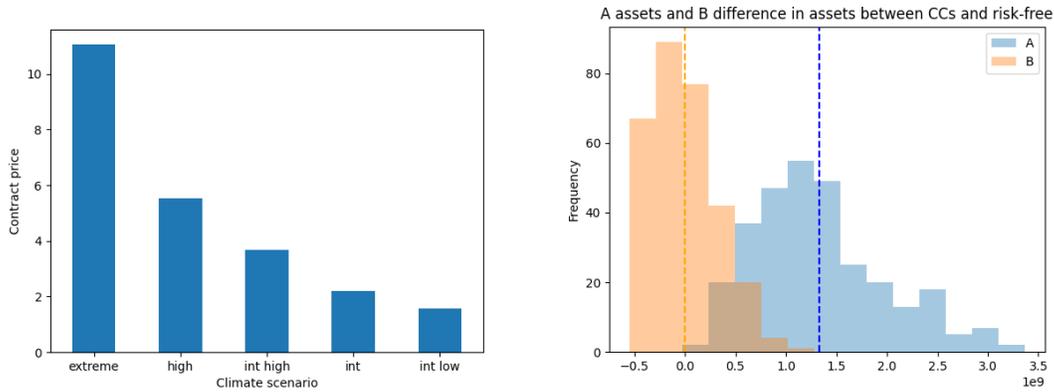

**Figure 20:** Prices (left chart) and distribution of outcomes for *A* and *B* with those prices (right chart).

*A* has a positive expected change in wealth and *B* has no expected change in wealth above risk-free investing. We can find optimal prices that make *A* and *B* as equally well off as possible by using machine learning to search for prices that minimize the difference between *A*'s and *B*'s expected value of assets, where the expected values are estimated by averaging the outcomes from replications of stochastic simulations.

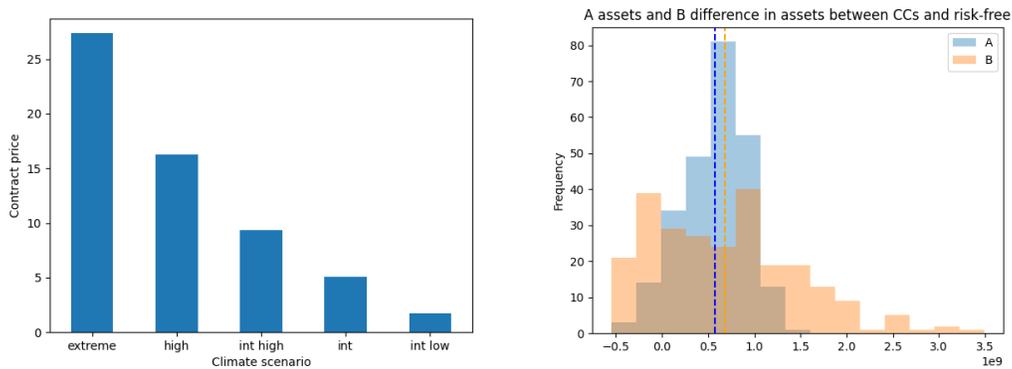

**Figure 21:** Prices found by optimization procedure (left) and distribution of outcomes with those prices (right).

With the prices the machine learning process discovered (Figure 21, left chart), the expected values for both parties are approximately equal (Figure 21, right chart). Given the adaptation payoff and discount factors, these prices offer the highest overall mutual gains to participants in CC transactions.[101] With these prices, the expected *annualized outperformance* over risk-free investing for the ***B*** investor is approximately 4.2% (a 5.2% annualized total return), which is very high for a hedging-focused investment product that is uncorrelated from other investment return streams, especially in the current financial environment where expected returns from traditional asset classes are at one of the lowest points in history.[102]

These instruments establishing one-time payment triggered by a climate threshold (a zero-coupon climate contingent bond) are not the only plausible implementation of climate contracts. Alternatively, ***A*** could issue coupon-paying bonds where the interest rate is contingent on the climate variable. After an initial period (e.g., 2 years), the interest rate could be tied to the climate variable. If the climate effects end up exactly as expected by the consensus climate model projection, then, if the bond had its rate linked to the climate variable, the repayment would look exactly like a traditional bond. If the climate is more extreme than expected, then building the adaptation proactively will have a higher expected rate of return and the bond interest rate will be higher. If the climate is less extreme than expected, then building the adaptation proactively will have a lower expected rate of return and the bond interest rate will be lower.

## VII. CLIMATE-CONTINGENT BONDS

Most climate adaptation is financed in the U.S. by municipalities issuing long-term bonds. Many bonds issued for financing real asset projects could eventually be climate-contingent bonds (CCBs).[103] Conceptually, a CCB is a blend of a climate contract and a traditional bond. It is a bond with a repayment rate connected to a climate change index, like mean sea-level rise, and provides a lower cost of capital to the issuer when the climate is less severe. Investors in CCBs are willing to accept below-market rates of return under less severe climate conditions because they would receive higher than market rates of return under more severe climate conditions. The returns are expected to be equal to a traditional bond issued at the market interest rate. Connecting the cost of capital to the climate scenarios that the financed projects are designed to address can unlock opportunities to proactively tackle the more extreme scenarios by, in effect, tying returns to the *A*'s ability to pay as a result of effectively generating value from their physical investments.

---

[101] If significant numbers of unaffiliated ***B*** investors bid competitively on the same contract, prices could be arbitraged lower from these "optimized" prices toward minimum acceptable prices based on risk-free investment returns and climate probabilities. The prices could even be bid down below the "minimum acceptable prices" if there were a significant number of ***B*** participants with non-financial environmental objectives. "According to a recent survey conducted by Morgan Stanley's Institute for Sustainable Investing, nearly 95% of millennials are interested in sustainable investing, while 75% believe that their investment decisions could impact climate change policy." *ESG Index Funds Hit $250 Billion as a Pandemic Accelerates Impact Investing Boom*, CNBC (Sept. 2, 2020, 9:25 AM), https://www.cnbc.com/2020/09/02/esg-index-funds-hit-250-billion-as-us-investor-role-in-boom-grows.html.

[102] Memorandum from Howard Marks to Oaktree Capital Management LP Clients (Oct. 13, 2020), https://www.oaktreecapital.com/docs/default-source/memos/coming-into-focus.pdf.

[103] Or, at least, be traditional bonds with a climate contract wrapper, which is effectively equivalent to a CCB.

Infrastructure projects in the planning phase can be built to defend against more extreme climate change through upfront spending that allows the city or company to reliably operate under many climate scenarios without having to undertake another future unplanned project expansion, and they generate a higher return on investment in more extreme climate scenarios. Meanwhile, for counterparties facing financial downside in the more extreme climate, it is optimal to invest because they can obtain higher than market rates of return exactly when they would need it the most from a climate risk perspective. Therefore, we should design CCBs to exhibit the following characteristics: the expected net present value is the same as a market rate bond of similar credit risk and lifetime term; if climate change is more extreme than expected, the interest rate is higher than market, but always less than some maximum interest rate; and if the climate change index is lower than expected, the interest rate is lower, but always greater than some minimum interest rate.

We use the example above where the U.S. Army Corps of Engineers calculated relative sea-level change (RSLC) projections in NYC for an analysis of flood risk mitigation infrastructure investment. They stated that the "use of the low or historic rate of RSLC will favor perimeter measures in plan selection, while the use of the high rate of sea level change favors larger barriers."[104] Given the significant uncertainty over the rate of sea level rise and that the city could be imperiled under high sea levels, the city could address this by financing larger barriers, infrastructure designed to withstand high sea-levels, with a CCB.

We can determine the CCB repayment structure with the following example input.

| Variable | Example city input |
|---|---|
| Climate Variable | Avg number of days per year of high-tide flooding in Northeast U.S., e.g., 100 |
| Lifetime of Bond | 25 years (2022-2047) |
| Discount Rate | 1% |
| Market Interest Rate | 4% |
| Minimum Interest Rate | 1% |
| Maximum Interest Rate | 7% |
| Granularity | 15 |

---

[104] U.S. ARMY CORPS OF ENG'RS, NEW YORK DIST., NEW YORK-NEW JERSEY HARBOR AND TRIBUTARIES COASTAL STORM RISK MANAGEMENT INTERIM REPORT Appendixes 8-10, (Feb. 2019), https://www.nan.usace.army.mil/Portals/37/docs/civilworks/projects/ny/coast/NYNJHAT/NYNJHAT Interim Report Economics Appendix Feb2019.pdf?ver=2019-02-19-165123-120.

**Table 2**: Inputs to determine a CCB repayment structure.

Software can then generate a distribution of possible outcomes of the Climate Variable from projections of climate variables derived from government data,[105] over Lifetime of Bond.

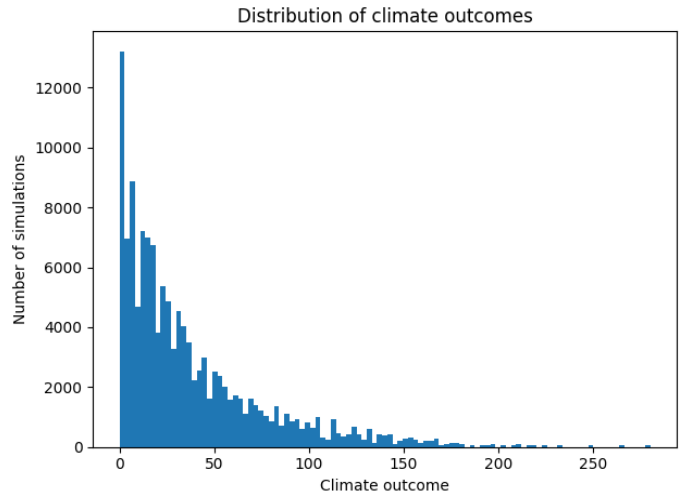

**Figure 22:** Simulated distribution of the number of high tide flooding days per year over the 2021-2046 period.[106]

Next, software can slice the climate outcome distribution into Granularity bins that are each equally likely to occur according to the climate distribution above. Then, the software finds the annual interest rate payment (the coupon rate) that NYC pays to the bond investors for the climate outcome on any given year. As a result, the expected net present value of the total return of the CCB would equal the net present value of the total return of the traditional bond. Software can determine the net present value of the total returns that a traditional bond at the Market Interest Rate would return over the Lifetime using the Discount Rate. The software can also estimate the expected total return of the CCB by setting a coupon rate for each of the climate outcome bins to a value – simulating thousands of possible future climate outcomes during the lifetime of the bond – and then taking the mean of the discounted CCB returns across those simulated time series. This average is the expected return.

The software runs this batch of simulations for a given set of coupon rates.[107] It then simulates and computes the expected return for another set of rates, using machine learning to intelligently search for rates and repeating this process thousands of times until it has converged on a set of rates where the expected total return is very close to the traditional bond. Figure 23 demonstrates the software's findings.

---

[105] Figure 22.
[106] Data from WILLIAM V. SWEET ET AL., NOAA TECHNICAL REPORT NOS CO-OPS 096: PATTERNS AND PROJECTIONS OF HIGH TIDE FLOODING ALONG THE U.S. COASTLINE USING A COMMON IMPACT THRESHOLD (2018), https://tidesandcurrents.noaa.gov/publications/techrpt86_PaP_of_HTFlooding.pdf.
[107] Figure 23 is an example of a set of coupon rates.

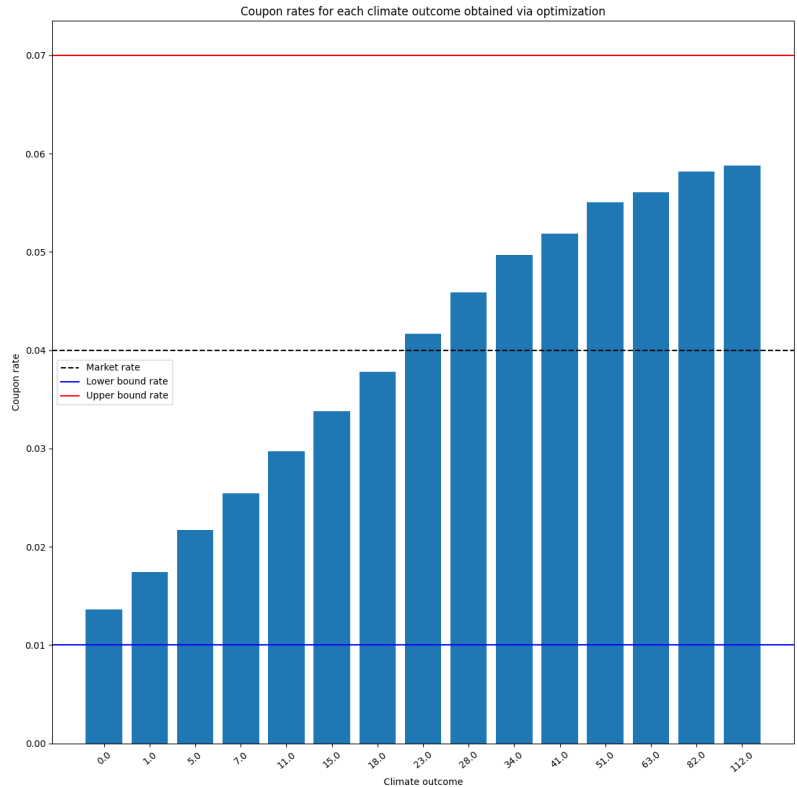

**Figure 23:** For each bin of possible climate outcome, the coupon rate to be paid to investors that year.

The labels on the climate outcomes in Figure 23 correspond to the bottom value of the climate bin. For instance, if there are 2 high tide flooding days on average across the government's measurement stations in the Northeast U.S. in a given year, the interest rate paid would correspond to the bin that has a "1.0" label. If there are 10 high-tide flooding days that year, the interest rate would correspond to the bin that has a "7.0" label.

Figures 24 and 25 illustrate 2,000 simulations of CCB repayment using the optimized rates. Each simulation samples from each year's potential climate outcome distribution, from 2022 to 2047, and NYC pays the investor according to the climate outcome realized that year.

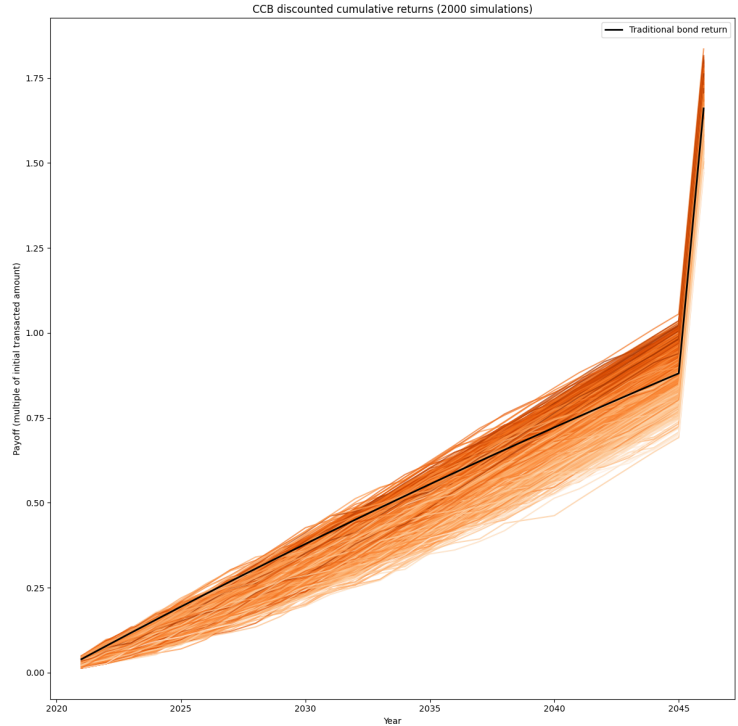

**Figure 24:** Cumulative return over time on a CCB (each colored line is a separate simulation) and a traditional bond (in black). Darker lines indicate a more severe climate occurred (on average).

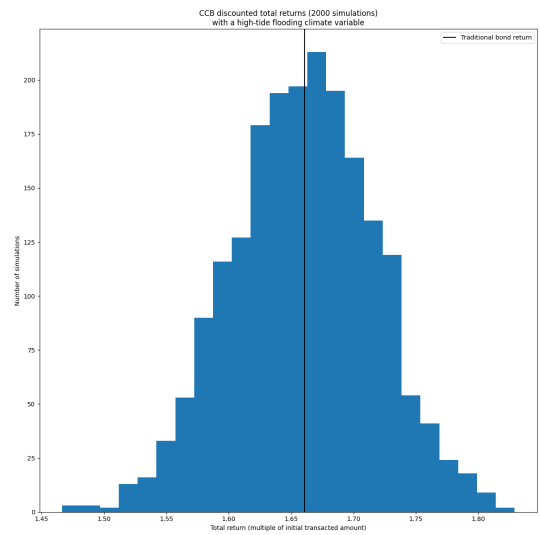

**Figure 25:** Total return across 2,000 simulations of the CCB with the coupon rates from the optimization process (histogram is counting the number of simulations with that total return) and a traditional bond at the market interest rate (vertical line).

## VIII. FEDERAL POLICY FOR CLIMATE-CONTINGENT FINANCE

Climate-contingent financial instruments can find natural places within the current regulatory regime. Climate contracts are derivative instruments dependent on a climate index. When the underlying index is economically meaningful to the contract's participants, the contract can likely be regulated as a "swap" contract exclusively by the U.S. Commodity Futures Trading Commission (CFTC) under the Commodity Exchange Act.[108] Climate-contingent bonds would likely fall under the purview of the Securities and Exchange Commission.[109]

The federal government currently incentivizes investments in public infrastructure by exempting income generated from interest paid by applicable municipal bond issuers from federal taxes. This indirectly reduces borrowing costs for municipalities because investors are willing to accept lower interest rates when the taxes on the interest they earn are lower, saving borrowers billions of dollars a year and incentivizing building more public infrastructure.[110]

To catalyze the climate-contingent market and incentivize forward-looking adaptation to climate change, the federal government could also exempt income generated from climate-contingent financial instruments issued by public (and possibly private) entities from federal taxes. As we have demonstrated in this Article, climate-contingent instruments make the counterparties better off in expectation. Furthermore, there is a positive externality to climate adaptation beyond the counterparties. Although not as large as the positive externality to climate mitigation (the reduction of greenhouse gases), climate adaptation in one location bolsters the resilience of the country as a whole. The exemption would not apply to tax-exempt investors such as non-profit endowments, but there is a large enough pool of tax-paying entities that this new exemption would be significant.

The federal government is the ultimate financial backstop for climate risk and, as a result, effectively holds trillions of dollars of climate risk on its balance sheet. The federal government could reduce its expected liabilities more than it loses in hypothetical tax dollars, producing a net positive financial outcome for taxpayers and increasing the resilience of the country, by incentivizing risk reduction through reduced taxation on income from climate-contingent instruments.

## IX. CONCLUSION

In this Article, we described how climate uncertainty reduces beneficial long-term financing and offered the solution of climate-contingent finance. We then examined the generalized structure of this mechanism, which applies to any situation where multiple entities are exposed to a shared long-term risk (e.g., climate change, or a natural pandemic), and one type of entity can take proactive actions to benefit from addressing the risk if it occurs (e.g., through innovating on crops that would do well under extreme climate change or a vaccination technology that would address particular viruses) with funding from another type of entity that seeks a targeted financial return to ameliorate the downside if the risk unfolds. Many entities may be in a position to take actions today that could help avoid ruin in unlikely (but plausible) scenarios, but cannot justify the capital expenditure in situations where those scenarios may not occur. The contingent-risk mechanism can fund efforts to address a variety of long-term risks to humanity that would otherwise lack traditional financing, including extreme climate change, large asteroids hitting the earth, and supervolcanic eruptions.

---

[108] 7 U.S.C. § 1(a)(47)(A)(ii-iii). For further discussion, see Sevren Gourley, *Funding Adaptation: Financing Resiliency Through Sea Level Derivatives*, ENV'T L. REV. SYNDICATE (2017), https://www.nyuelj.org/2017/04/%EF%BB%BFfunding-adaptation-financing-resiliency-through-sea-level-derivatives/.

[109] 15 U.S.C. § 78c(a)(29); *see* Gourley, *supra* note 108.

[110] *See There's A Better Way to Pay for Infrastructure*, BLOOMBERG OPINION (May 18, 2021), https://www.bloomberg.com/opinion/articles/2021-05-18/there-s-a-better-way-to-pay-for-infrastructure?srnd=premium.

We investigated the specifics of how the financial mechanism applies to climate change. The Article explored this through case studies of cities (Ho Chi Minh City, Vietnam, Bristol, UK, and New York, USA), and a discussion of the various types of potential participating parties to climate contracts. We conducted extensive computational simulation experiments to explore simple climate-contingent contracts and more complex climate-contingent bonds. Optimization analyses of the simulation models illustrated how parameters of these financial instruments could be set for specific counterparties with specific climate data. Finally, the Article outlined a proposal to catalyze climate-contingent finance by exempting the income generated from federal taxes.

Governments, asset owners, and companies reduce uncertainty in components of the economy (e.g., commodities prices, credit risks, and interest rates) through trillions of dollars of derivatives positions and insurance contracts – climate-contingent instruments may provide a similar purpose in the climate context. Municipalities raise trillions of dollars of debt for infrastructure – climate-contingent instruments may allow those investments to align with climate-aware designs and, therefore, lead to a more climate-secure future.

Climate-contingent finance is a fresh approach to addressing catastrophic risk. It builds a bridge between long-term funding needs and financial risk management. An important next step in this research area is developing more in-depth, on-the-ground case studies – alongside local governments and potential long-term investors – for funding climate change adaptation infrastructure projects (that have been planned, but currently lack funding) with climate-contingent bonds. The methods developed in this Article are also primed for research applications within other long-term risk areas, such as natural pandemics and supervolcanic eruptions. With further research into risk-contingent financing, the massive fixed-income markets can be a source of profound positive impact toward safeguarding humanity's long-term flourishing.

# APPENDIX A: CLIMATE RISK PRICING

We have developed a conceptual framework for estimating the speed and magnitude of the pricing-in of climate risk. Theoretically, it could reflect expectations of the impacts of climate transition risk and climate physical risk into asset prices. Climate physical risk is materially increasing on a decadal timeline. Climate transition risk — law and policy that attempts to reduce the concentration of greenhouse gases in the atmosphere, low-carbon technology development and deployment, and consumer sentiment changes — is materially increasing on a monthly and annual timeline.

There are three key drivers of climate physical risk being reflected in prices:[111]

1. The median holding period (and, by implication, the time horizon) of the average asset owner (e.g., publicly traded equity shares have significantly shorter holding periods than primary residences);
2. The expected value, and the potential for fat tails in the distribution of possible outcomes, of the hazard's impact on future cash-flow in a way that is not easily addressed by low-cost climate adaptation;
3. The extent to which the average asset owner can diversify away the risk (e.g., primary residences are the lowest on this dimension and public equities are the highest).

If 1 and 2 are high, and 3 is low, we would expect that prices for that type of asset to reflect climate expectations the most. A driving macro factor across all asset classes is the real interest rate: when it decreases, the discount factor applied to future cash flows decreases and therefore the time horizon for long-term impacts on all assets is increased.

The more sensitive the climate is to the level of emissions, the more both physical and transition risks are amplified because there will be a greater need to reduce emissions and more associated physical risk. High uncertainty around emissions-climate sensitivity increases the ambiguity of dimension 2 because there are fatter tails in the distribution of potential climate impact on future free cash-flows and in the distribution of potential efficacy of physical adaptation investments. The more transition risk, all else equal, the less physical risk because emissions will be further reduced by the aggressive transition.

For physical changes, the most net impactful climate hazard types for non-sovereign asset owners are chronic (e.g., sea-level rise), rather than acute (e.g., hurricanes). We use "net" impactful because intergovernmental transfers are much more likely around acute impacts relative to chronic impacts. Acute disaster events often trigger (inter-)national disaster declarations and the subsequent release of (inter-)national funds for aid and rebuilding, reducing the local entity or asset owner's ultimate cost. In the U.S., this phenomenon is colloquially referred to as the "FEMA put,"

---

[111] Less of a government backstop, and more liquidity, are two additional factors that increase the price sensitivity to investor expectations. These are generally applicable factors, not climate specific.

where the Federal Emergency Management Agency serves as a financial backstop for acute disaster situations when/where the President declares an official emergency.

# APPENDIX B: SEA-LEVEL RISE IMPACTS, AND ADAPTATION OPTIONS

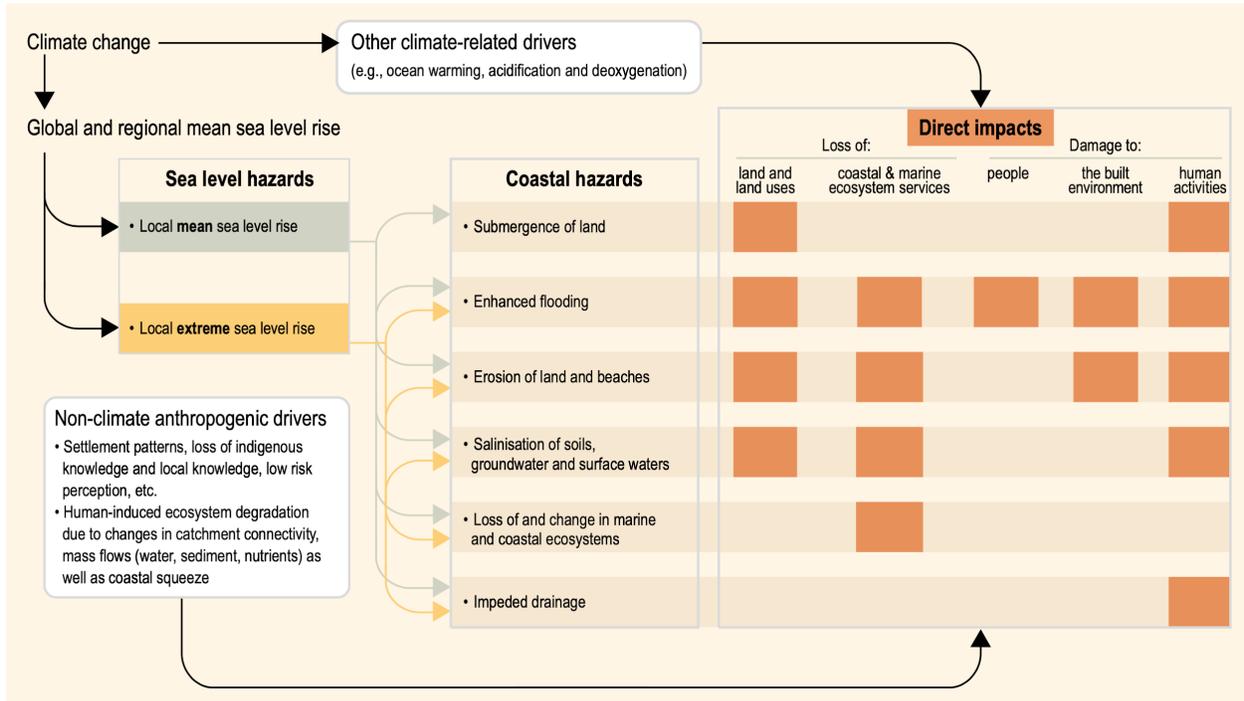

112

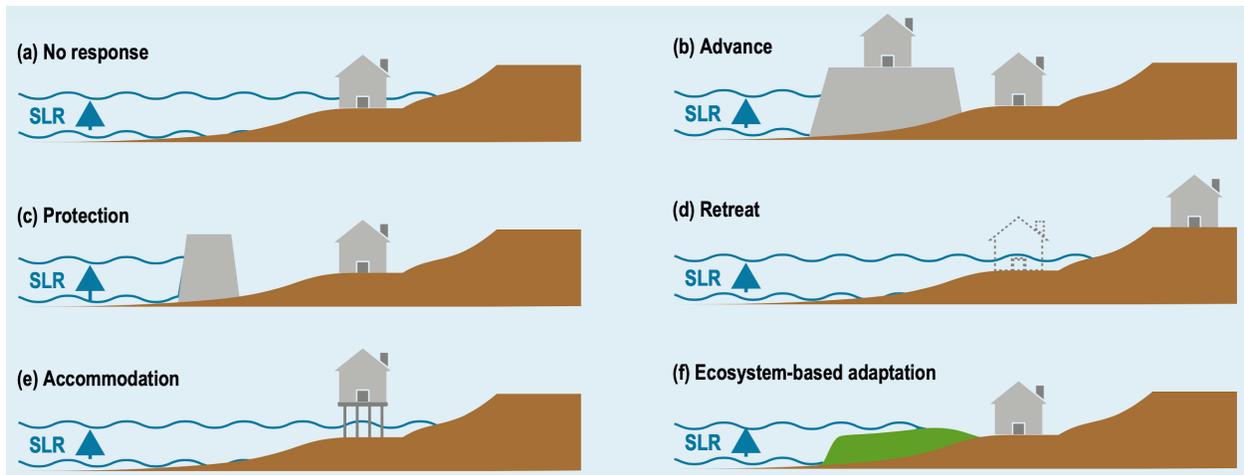

113

---

[112] Figure from OPPENHEIMER ET AL., *supra* note 58, at 375.
[113] Figure from *id.* at 386.